\pdfoutput=1

\documentclass[11pt,twoside,a4paper,cmspaper,final,collab]{cms-tdr}
\def\svnVersion{390806}\def\cmsCernNoTag{CERN-EP/2016-208}\def\cmsCernDate{\today}\def\cmsMessage{Published in the Journal of High Energy Physics as \href{http://dx.doi.org/10.1007/JHEP02(2017)079}{\doi{10.1007/JHEP02(2017)079.}}}

\begin{document}\cmsNoteHeader{TOP-13-017}

\hyphenation{had-ron-i-za-tion}
\hyphenation{cal-or-i-me-ter}
\hyphenation{de-vices}
\RCS$HeadURL: svn+ssh://svn.cern.ch/reps/tdr2/papers/TOP-13-017/trunk/TOP-13-017.tex $
\RCS$Id: TOP-13-017.tex 381703 2017-01-19 22:10:34Z alverson $

\newlength\cmsFigWidth
\ifthenelse{\boolean{cms@external}}{\setlength\cmsFigWidth{0.85\columnwidth}}{\setlength\cmsFigWidth{0.4\textwidth}}
\ifthenelse{\boolean{cms@external}}{\providecommand{\cmsLeft}{top}}{\providecommand{\cmsLeft}{left}}
\ifthenelse{\boolean{cms@external}}{\providecommand{\cmsRight}{bottom}}{\providecommand{\cmsRight}{right}}
\newcommand{\x}{\ensuremath{\phantom{0}}}

\newcommand{\WHIZARD} {\textsc{whizard}\xspace}
\providecommand{\rj}{\ensuremath{\mathrm{j}}}
\hyphenation{had-ron-i-za-tion}
\hyphenation{cal-or-i-me-ter}
\hyphenation{de-vices}

\cmsNoteHeader{TOP-13-017}

\title{
    Search for top quark decays via Higgs-boson-mediated flavor-changing
    neutral currents in pp collisions at $\sqrt{s} = 8\TeV $
}

\date{\today}
\abstract{A search is performed for Higgs-boson-mediated flavor-changing
          neutral currents in the decays of top quarks.  The search is based
          on proton-proton collision data corresponding to an integrated
          luminosity of 19.7\fbinv at a center-of-mass energy of 8\TeV
          collected with the CMS detector at the LHC.  Events in which a top
          quark pair is produced with one top quark decaying into a charm or
          up quark and a Higgs boson (\PH), and the other top quark decaying
          into a bottom quark and a \PW boson are selected.  The Higgs boson in
          these events is assumed to subsequently decay into either dibosons
          or difermions.  No significant excess is observed above the expected
          standard model background, and an upper limit at the 95\% confidence
          level is set on the branching fraction $\mathcal{B}\left({\rm t \to
          Hc}\right)$ of 0.40\% and $\mathcal{B}\left({\rm t \to Hu}\right)$
          of 0.55\%, where the expected upper limits are 0.43\% and 0.40\%,
          respectively.  These results correspond to upper limits on the
          square of the flavor-changing Higgs boson Yukawa couplings $|\lambda_{\rm
          tc}^{\mathrm{H}}|^{2} < 6.9\times 10^{-3}$ and $|\lambda_{\rm tu}^{\mathrm{H}}|^{2} <
          9.8 \times 10^{-3}$.
         }
\hypersetup{
pdfauthor={CMS Collaboration},
pdftitle={
    Search for top quark decays via Higgs-boson-mediated flavor-changing
    neutral currents in pp Collisions at sqrt(s) = 8 TeV},
    pdfsubject={top}, pdfkeywords={CMS, physics, Higgs, top, searches}
}
\maketitle
\section{Introduction}
\label{sec:introduction}

\par
With the discovery of the Higgs boson (\PH)~\cite{higgs:atlas,higgs:cms,long}
it is possible to probe new physics by measuring its coupling to
other particles.  Of particular interest is the flavor-changing neutral
current (FCNC) decay of the top quark to the Higgs boson. The
investigation of this process at the CERN LHC is motivated by the large
\ttbar production cross section and the variety of possible decay modes
of the Higgs boson.

The next-to-next-to-leading-order $\ttbar$ production cross section at a
center-of-mass energy of 8\TeV and with a top quark mass ($m_{\rm t}$)
of 173.5\GeV~\cite{PDG} is 252\unit{pb}~\cite{PhysRevLett.110.252004}. The standard
model~(SM) predicts that the top quark decays with a branching fraction
of nearly 100\% into a bottom quark and a \PW boson ($\rm t \to
Wb$).

In the SM, FCNC decays are absent at leading-order and occur only via
loop-level processes that are additionally suppressed by the
Glashow-Iliopoulos-Maiani mechanism \cite{Eilam:1990zc,sm-pred-thc2}.
Because the leading-order decay rate of $\rm t \to Wb$ is also quite
large, the SM branching fraction $\mathcal{B}\left({\rm t \to
Hq}\right)$, where q is an up or charm quark, is predicted to be of
$\mathcal{O}(10^{-15})$~\cite{Eilam:1990zc,sm-pred-thc2,sm-pred-thc3},
far below the experimental sensitivity at the LHC.  However, some
extensions of the SM predict an enhanced $\rm t \to Hq$ decay rate.
Thus, observation of a large branching fraction would be clear
evidence for new physics.  The largest enhancement in $\mathcal{B}({\rm
t \to Hq})$ is predicted in models that incorporate a two-Higgs doublet,
where the branching fraction can be of
$\mathcal{O}(10^{-3})$~\cite{sm-pred-thc3}.

Previous searches for FCNC in top quark decays mediated by a Higgs boson
have been performed at the LHC by ATLAS~\cite{Aad:2014dya,Aad:2015pja} and
CMS~\cite{Khachatryan:2014jya}.  The CMS search considered both
multilepton and diphoton final states and the observed upper limit of
$\mathcal{B}({\rm t \to  Hc})$ at the 95\% confidence level (CL) was
determined to be 0.56\%.  The recent ATLAS result included final states
where the Higgs boson decays to b quark pairs, and measured the observed
upper limits of $\mathcal{B}({\rm t \to  Hc})$ and $\mathcal{B}({\rm t
\to  Hu})$ at the 95\% CL to be 0.46\% and 0.45\%, respectively.

The analysis presented here uses a data sample recorded with the CMS
detector and corresponding to an integrated luminosity of 19.7\fbinv of
\Pp\Pp\, collisions at $\sqrt{s} = 8\TeV$.  The data were recorded in
2012 with instantaneous luminosities  of  5--8\,$\times10^{33}~\rm
cm^{-2}s^{-1}$ and an average of 21 interactions per bunch crossing. The
inelastic collisions that occur in addition to the hard-scattering
process in the same beam crossing produce mainly low-\pt particles that
form the so-called ``pileup'' background.

In this paper, the FCNC decays $\cPqt\to\PH\cPqc$ and $\cPqt\to\PH\cPqu$
are searched for through the processes $\ttbar \to \rm Hc+Wb$ or $\rm
Hu+Wb$.  Three independent analyses are perfomed and their results are
then combined.  The multilepton analysis considers events with two
same-sign (SS) leptons or three charged leptons (electrons or muons).
This channel is sensitive to the Higgs boson decaying into WW, ZZ, or
$\tau\tau$ which have branching fractions of 21.5\%, 2.6\%, and 6.3\%,
respectively~\cite{YR3}.  The diphoton analysis considers events with
two photons, a bottom quark, and a W boson that decays either
hadronically or leptonically.  The two photons in this channel are used
to reconstruct the Higgs boson which decays to diphotons with
$\mathcal{B}\left({\rm H}\to \gamma\gamma \right) =
0.23\%$~\cite{YR3}.  Finally, events with at least four jets, three of
which result from the hadronization of bottom quarks (b jets), and a
leptonically decaying W boson are considered.  The b jet + lepton
channel takes advantage of the large Higgs boson branching fraction into
\bbbar pairs, $\mathcal{B}(\PH\to\bbbar) =
57\%$~\cite{Denner:2011mq}.  A summary of the enumerated final states is
shown in Table~\ref{topologies}.

The CMS detector and trigger are described in Section~\ref{sec:cms}, and
the event selection and reconstruction in
Section~\ref{sec:preselection}.  Section~\ref{sec:data} then discusses
the Monte Carlo (MC) simulation samples.  The signal selection and
background estimations for each of the three analyses are given in
Section~\ref{sec:signal}, and the systematic uncertainties in
Section~\ref{sec:sys}.  Finally, the individual and combined results
from the analyses are presented in Section~\ref{sec:results}.

\begin{table}[ht]
    \begin{center}
    \topcaption{
        Summary of the requirements for the ${\rm pp \to \ttbar \to Hq + Wb}$ channels used
        in this analysis.
        \label{topologies}}

    \resizebox{\textwidth}{!}{
    \begin{tabular}{ l | c  c  c  c  c }
    \multicolumn{1}{c}{Decay channels} & Leptons & Photons & Jets & b jets & Category \\
    \hline
    H $\to$ WW, ZZ, $\tau\tau$  \& W$\to\ell\nu$ & eee, ee$\mu$, e$\mu\mu$, $\mu\mu\mu$ & --- & ${\ge}2$ & --- & trilepton \\
    H $\to$ WW, ZZ, $\tau\tau$  \& W$\to\ell\nu$  & e$^\pm$e$^\pm$, e$^\pm\mu^\pm$, $\mu^\pm\mu^\pm$ & --- & ${\ge}2$ & --- & dilepton SS \\
    H $\to$ $\gamma\gamma$ \& W$\to\ell\nu$ & e$^\pm$, $\mu^\pm$ &  ${\ge}2$ &  ${\ge}2$ & =1 & diphoton + lepton  \\
    H $\to$ $\gamma\gamma$ \& W$\to q_1 q_2$ & ---                &  ${\ge}2$ &  ${\ge}4$ & =1 & diphoton + hadron  \\
    H $\to$ $\bbbar$ \& W$\to\ell\nu$ & e$^\pm$, $\mu^\pm$ &  --- & ${\ge}4$ &  ${\ge}3$ & b jet + lepton \\
    \end{tabular}
    }
    \end{center}
\end{table}

\section{The CMS detector and trigger}
\label{sec:cms}

A detailed description of the CMS detector, together with a definition
of the coordinate system used and the relevant kinematic variables, can
be found in Ref.~\cite{ref:cms}.  The central feature of the CMS
apparatus is a superconducting solenoid, 13\unit{m} in length and
6\unit{m} in diameter, which provides an axial magnetic field of
3.8\unit{T}.  Within the field volume there are several particle
detection systems. Charged particle trajectories are measured by silicon
pixel and strip trackers, covering $0\le \phi \le 2\pi$ in azimuth and
$\abs{\eta} < 2.5$ in pseudorapidity. A lead tungstate crystal
electromagnetic calorimeter (ECAL) surrounds the tracking volume. It is
comprised of a barrel region $\abs{\eta} < 1.48$ and two endcaps that
extend up to $\abs{\eta} = 3$.  A brass and scintillator hadron
calorimeter (HCAL) surrounds the ECAL and also covers the region
$\abs{\eta} < 3$.  The forward hadron calorimeter (HF) uses steel as the
absorber and quartz fibers as the sensitive material. The HF extends the
calorimeter coverage to the range $3.0 < |\eta| < 5.2$.  A lead and
silicon-strip preshower detector is located in front of the ECAL
endcaps. Muons are identified and measured in gas-ionization detectors
embedded in the steel flux-return yoke outside the solenoid.  The
detector is nearly hermetic, allowing momentum balance measurements in
the plane transverse to the beam direction.

Depending on the final state under consideration, events are selected at
the trigger level by either requiring at least two leptons, ($\Pe\Pe$,
$\mu\mu$ or $\Pe\mu$), at least two photons, or a single lepton ($\Pe$
or $\mu$) to be within the detector acceptance and to pass loose
identification and kinematic requirements.

The dilepton triggers used in the multilepton selection require one
lepton with $\pt > 17\GeV$ and one lepton with $\pt > 8\GeV$.  At
the trigger level and during the offline selection, electrons are required to
be within $\abs{\eta} < 2.5$, and muons are required to be within
$\abs{\eta} < 2.4$.  All leptons must be isolated, as described in
Section~\ref{sec:preselection}, and have $\pt > 20\GeV$ for the
highest-\pt lepton, and $\pt > 10\GeV$ for all subsequent leptons in the
event.  For events satisfying the full multilepton selection, the dimuon,
dielectron, and electron-muon trigger efficiencies are measured to be
98\%, 91\%, and 94\%, respectively, for the SS dilepton selection, and
100\% for the trilepton selection.

The diphoton trigger requires the presence of one photon with $\pt >
36\GeV$ and a second photon with $\pt > 22\GeV$.  Loose isolation and
shower shape requirements are applied to both photons~\cite{EGM-14-001}.
The average diphoton trigger efficiency is measured to be 99.4\% after
applying the full event selection for photons within $\abs{\eta} < 2.5$,
excluding the barrel-endcap transition region $1.44 <\abs{\eta} < 1.57$.

The b jet + lepton selection uses the single-lepton triggers.  The
single-muon trigger requires at least one isolated muon with $\pt >
24\GeV$ and $\abs{\eta} < 2.1$ to be reconstructed online.  The
single-electron trigger requires at least one isolated electron with
$\pt > 27\GeV$  and $\abs{\eta} < 2.5$.  The offline selection further
requires that electrons have $\pt > 30\GeV$ and muons have $\pt >
26\GeV$.  This results in an average trigger efficiency of 84\% for the
single-electron triggers and 92\% for the single-muon trigger after the
b jet + lepton selection.

\section{Event selection and reconstruction}
\label{sec:preselection}

Events are required to have a primary vertex with a reconstructed
longitudinal position within 24\unit{cm} of the geometric center of the
detector and a transverse position within 2\unit{cm} from the nominal
interaction point.  To distinguish the hard-scattering vertex from
vertices arising from pileup interactions, the reconstructed vertex with
the highest scalar sum of the ${\pt^2}$ of its associated tracks is
chosen as the primary vertex.  To ensure that leptons originate from
the same primary vertex, a loose requirement is applied to their
longitudinal and transverse impact parameters with respect to the
primary vertex.

The particle-flow event algorithm~\cite{ref:pf,CMS:2010byl} is used to
reconstruct and identify individual particles using an optimized
combination of information from the elements of the detector.  Prompt
electrons and muons arising from W and Z decays are typically more
isolated than nonprompt leptons arising from the decay of hadrons within
jets.  In order to distinguish between prompt and nonprompt lepton
candidates, a relative isolation parameter is defined for each lepton
candidate.  This is calculated by summing the \pt of all charged and
neutral particles reconstructed using the particle-flow algorithm within
a cone of angular radius $\DR \equiv \sqrt{\smash[b]{(\Delta\eta)^2 +
(\Delta \phi)^2}} = 0.4$ around the lepton candidate momentum, where
$\Delta\eta$ and $\Delta\phi$ are the pseudorapidity and azimuthal angle
(in radians) differences, respectively, between the directions of the
lepton and the other particle~\cite{Baffioni:2006cd,Chatrchyan:2012xi}.
This cone excludes the lepton candidate and the charged particles
associated with the pileup vertices.  The resulting quantity is
corrected for additional underlying-event activity owing to neutral
particles~\cite{long}, and then divided by the lepton candidate's \pt.
The relative isolation parameter is required to be less than
0.15 for electrons and 0.12 for muons.

The electron selection criteria are optimized using a multivariate
approach that combined information from both the tracks and ECAL clusters,
and have a combined identification and isolation efficiency of
approximately 60\% at low \pt (10\GeV) and 90\% at high \pt (50\GeV) for
electrons from $\PW$ or $\Z$ boson decays~\cite{Khachatryan:2015hwa}.
The training of the multivariate electron reconstruction is performed
using simulated events, while the performance is validated using data.

Muon candidates are reconstructed with a global trajectory fit using
hits in the tracker and the muon system. The efficiency for muons to
pass both the identification and isolation criteria is measured from
data to be larger than 95\%~\cite{eleReg1b,long}.

For events in which there is an overlap between a muon and an electron,
\ie, an electron within $\DR < 0.1$ of a muon, precedence is given to
the muon by vetoing the electron.  In the multilepton selection, events
in which there are more than three isolated leptons (electron or muon)
with \pt $>$ 10\GeV are rejected to reduce diboson contamination.  The
invariant mass of dilepton pairs in the SS channel is required to be
greater than 30\GeV in order to reject low-mass resonances and reduce
poorly modeled backgrounds (\eg, QCD).  In the b jet + lepton selection,
events in which there are additional isolated electrons with
$\pt > 20\GeV$ and $\abs{\eta} < 2.5$ or isolated muons with
$\pt > 10\GeV$ and $\abs{\eta} < 2.4$ are rejected.

The photon energy is reconstructed from the sum of signals in the ECAL
crystals~\cite{EGM-14-001}.  The ECAL signals are
calibrated~\cite{Calib-ECAL}, and a multivariate regression, developed
for a previous $\PH\to\gamma\gamma$ analysis~\cite{cms-Hgg-Legacy}, is
used to estimate the energy of the photon.  Clusters are
formed from the neighboring ECAL crystals seeded around local maxima
of energy deposits, and the collection of clusters that contain the
energy of a photon or an electron is called a supercluster.
Identification criteria are applied to distinguish photons from jets
and electrons.  The observables used in the photon identification
criteria are the isolation variables, the ratio of the energy in the
HCAL towers behind the supercluster to the electromagnetic energy in
the supercluster, the transverse width in $\eta$ of the
electromagnetic shower, and the number of charged tracks matched to
the supercluster.  The photon efficiency identification is measured
using $\Z \to \Pe^{+}\Pe^{-}$ events in data by reconstructing the electron
showers as photons~\cite{CMS:2011aa}, taking into account the shower
shape and whether the electron probe is located in the barrel or
endcap.  The two highest \pt photons must exceed 33
and 25\GeV, respectively.

Jets are reconstructed from the candidates produced by the particle-flow
algorithm.  An anti-$\kt$ clustering algorithm~\cite{ref:kt} with a distance
parameter of 0.5 is used for jet reconstruction.  Jets with a significant
fraction of energy coming from pileup interactions or not associated with the
primary vertex are rejected.  Remaining pileup energy in jets is
subtracted using a technique that relies on information about the jet
area~\cite{pile1,pile2,pile3}.  Reconstructed jets are calibrated to
take into account differences in detector response~\cite{ref:jetscale}.  The
jets in the multilepton and b jet + lepton selections are required to have $\pt
> 30\GeV$, $\abs{\eta} < 2.5$, and to be separated from leptons such that
$\DR(\text{lepton, jet}) > 0.3$.  The selection of jets in the diphoton events
differs by requiring the jet $\ET > 20\GeV$ and the jets be
separated from both photons such that $\DR(\text{photon, jet}) > 0.3$.

To characterize the amount of hadronic activity in an event, the scalar sum of
the transverse energy of jets passing all of these requirements ($H_\mathrm{T}$)
is calculated.  The missing transverse energy (\MET) is calculated as the
magnitude of the vector sum of the transverse momenta of all reconstructed
particle-flow candidates in the event.

Jets originating from the hadronization of b quarks are identified by the
combined secondary vertex (CSV) b tagging algorithm~\cite{CSV}.  The selection
criteria that are used have an identification efficiency of 66\%, and a
misidentification rate of 18\% for charm quarks and 1\% for light-quark and
gluon jets.  The diphoton and b jet + lepton selections require
b-tagged jets.  Although the identification of b jets is not used to select
signal events in the multilepton selection, it is used for the purpose of
defining control samples to check the normalization of simulated background
processes.  No additional tagging is used to discriminate between jets
originating from c quarks.

The inclusion of b jets in the diphoton and b jet + lepton selections
results in a difference in the sensitivity to the $ \PQt \to \PH \PQu $ and
$ \PQt \to \PH \PQc $ decay modes.  This is caused by the larger likelihood
of b tagging a jet originating from a charm quark than from an up quark.
The multilepton analyses do not include b tagging to enhance the signal
sensitivity so the two FCNC top quark decay modes are indistinguishable.

\section{Simulated samples}
\label{sec:data}

The determination of the expected signal and background yields relies on
simulated events, as well as an estimation based on control samples in
data, as discussed in later sections. Samples of Drell--Yan, $\ttbar$,
\PW+jets, $\PW + \bbbar$, diboson, $\ttbar + \PZ$, $\ttbar + \PW$,
and triboson events are generated using the \MADGRAPH event generator
(v5.1.5.11)~\cite{Alwall:2011uj}.  The samples of $\PZ\PZ$ to four charged
leptons and single top quark events are generated using \POWHEG (v1.0
r1380)~\cite{ref:Nason:2004rx,ref:Frixione:2007vw,ref:Alioli:2010xd}.
In all cases, hadronization and showering are done through \PYTHIA
(v6.426)~\cite{ref:pythia}, and $\tau$ decays are simulated using
\TAUOLA (v2.75)~\cite{Was:2000st}.  Three additional production
processes are considered for the nonresonant diphoton backgrounds, where
the dominant one coming from $\gamma\gamma$ + jets is simulated with
\SHERPA (v1.4.2)~\cite{SHERPA}. Top quark pairs  with one additional
photon are simulated with \MADGRAPH, while those with two additional
photons are simulated using the \WHIZARD (v2.1.1)~\cite{WHIZARD}
generator interfaced with \PYTHIA.  The Z2 tune~\cite{Field:2011iq} of
\PYTHIA is used to model the underlying event.

Events that arise from the SM Higgs boson production are treated as a
background.  The gluon-fusion (ggH) and vector-boson-fusion (VBF) Higgs
boson production processes are generated with \POWHEG at
next-to-leading order (NLO) in QCD, interfaced with \PYTHIA .  The
associated W/ZH production and $\ttbar$H processes are simulated with
\PYTHIA at leading order. The cross sections and branching fractions of
the SM Higgs boson processes are set to the values recommended by the
LHC Higgs cross section working group~\cite{YR3}.

The simulated samples for the signal process ${\ttbar \to \rm Hq
+ Wb}$ (q = c or u) are produced using \PYTHIA for the case of the
Higgs boson decaying to WW, ZZ, $\tau\tau$, and $\gamma\gamma$, and with
\MADGRAPH for $\rm H \to \bbbar$.  The use of different generators is an
artifact of the various modes being analyzed separately.  The Higgs boson
is assumed to have a mass of 125\GeV.

The set of parton distribution functions (PDF) used is
CTEQ6L~\cite{ref:CTEQ6LL} in all cases, except for $\rm H \to
\bbbar$, where CT10~\cite{PhysRevD.78.013004} is used.

The CMS detector response is simulated using a $\GEANTfour$-based
(v9.4)~\cite{ref:geant} model, and the events are reconstructed and
analyzed using the same software used to process collision data.
The effect of pileup is included in the simulation process by
superimposing simulated events on the process of interest.  The
simulated signal events are weighted to account for the differences
between data and simulation of the trigger, reconstruction, and
isolation efficiencies, and the distributions of the reconstructed
vertices coming from pileup.  Additional corrections are applied to
account for the energy scale and lepton \pt resolution.  The observed jet
energy resolution and scale~\cite{ref:jetscale}, top quark
\pt distribution~\cite{CMS-PAS-TOP-12-027}, and b tagging
efficiency and discriminator distribution~\cite{Khachatryan:2014qaa} in
data are used to correct the simulated events.  Corrections accounting
for the differences in lepton selection efficiencies are derived using the
tag-and-probe technique~\cite{Chatrchyan:2011cm}.

\section{Signal selection and background estimation}
\label{sec:signal}

The sensitivity of the search is enhanced by combining the twelve exclusive
channels, shown in Table~\ref{topologies}, defined according to the
expected decay modes of the Higgs and W bosons.

\subsection{Multilepton channels}

\par
The multilepton analysis is conducted with the goal of
enhancing the signal sensitivity in the trilepton channel:
$\ttbar \to \PH\cPq + \PW\cPqb \to \ell\nu\ell\nu\cPq + \ell\nu\cPqb$,
and the SS dilepton channel: $\ttbar \to \PH\cPq + \PW\cPqb \to \ell\nu {
\cPq\cPq\cPq} + \ell\nu\cPqb$, where $\ell$ represents either a muon or
electron.  The main target of optimization is final states
resulting from $\rm H \to WW$ decays.

\begin{figure}[h]
    \begin{center}
    \vspace{1cm}
    \includegraphics[width=0.99\textwidth]{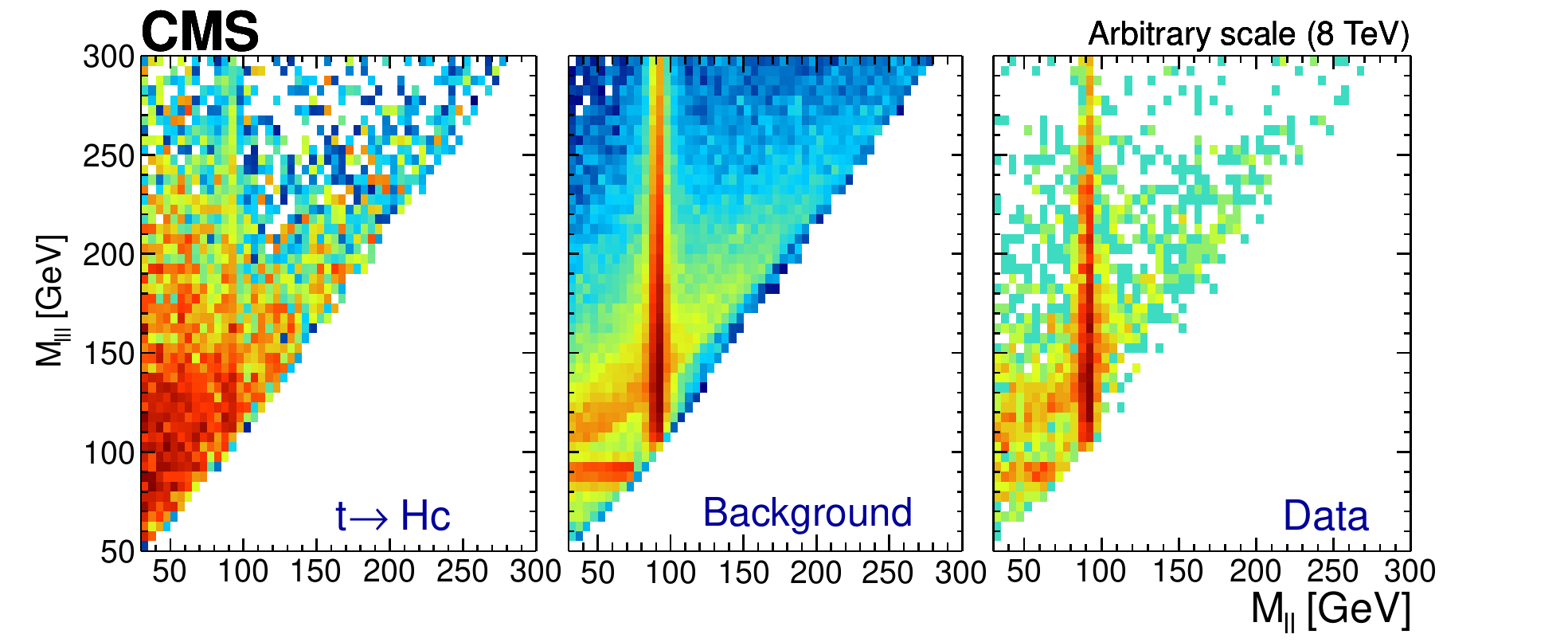}\\
    \caption{\label{fig:1}
    Trilepton invariant mass versus opposite-sign dilepton invariant
    mass in the trilepton channel after the event selection described in
    Section~\ref{sec:preselection} for simulated signal, estimated
    background, and data, from left to right.
    }
    \end{center}
\end{figure}

In the case of the trilepton channel, rejection of events containing
dileptons  originating from resonant \Z boson production is necessary to
remove backgrounds from WZ production, asymmetric internal conversions
(AIC, the process in which final-state radiation in a Drell--Yan event
converts to dileptons where one of the leptons carries most of the
photon momentum)~\cite{asy-conv} or final-state radiation where the
photon is misidentified as an electron.  A comparison of the
two-dimensional distribution of the trilepton mass versus the opposite-sign
dilepton mass is shown in Figure~\ref{fig:1} for the estimated signal
and background processes, and data.  Events satisfying any of
the following criteria are vetoed to reduce the contribution from
resonant \PZ production: (1)  the invariant mass of an opposite-sign,
same-flavor (OSSF) lepton pair is within 15\GeV of the \PZ boson
mass~\cite{PDG}; (2) the invariant mass of an OSSF lepton pair is
greater than 30\GeV and the trilepton invariant mass is within 10\GeV of
the \PZ boson mass.  For the SS dielectron channel, electron pairs with
an invariant mass within 15\GeV of the \PZ boson mass are rejected to
reduce the background arising from misidentification of the electron
charge.  No invariant mass requirement is applied to the
$\mu^{\pm}\mu^{\pm}$ and e$^{\pm}\mu^{\pm}$ final states since there is
a negligible contamination from resonant \PZ boson production.

\begin{figure}[h]
    \begin{center}
    \includegraphics[width=0.45\textwidth]{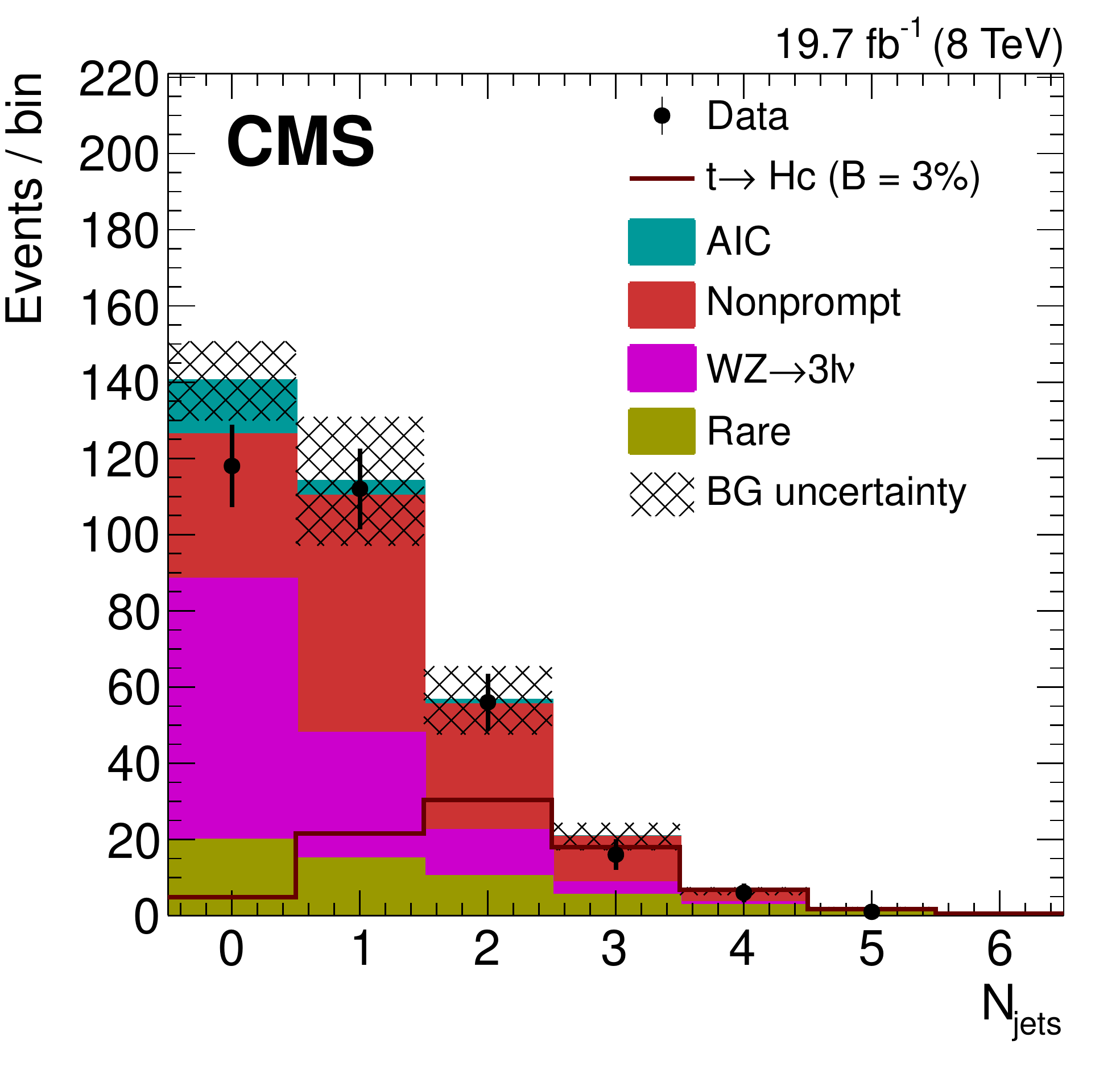}
    \includegraphics[width=0.45\textwidth]{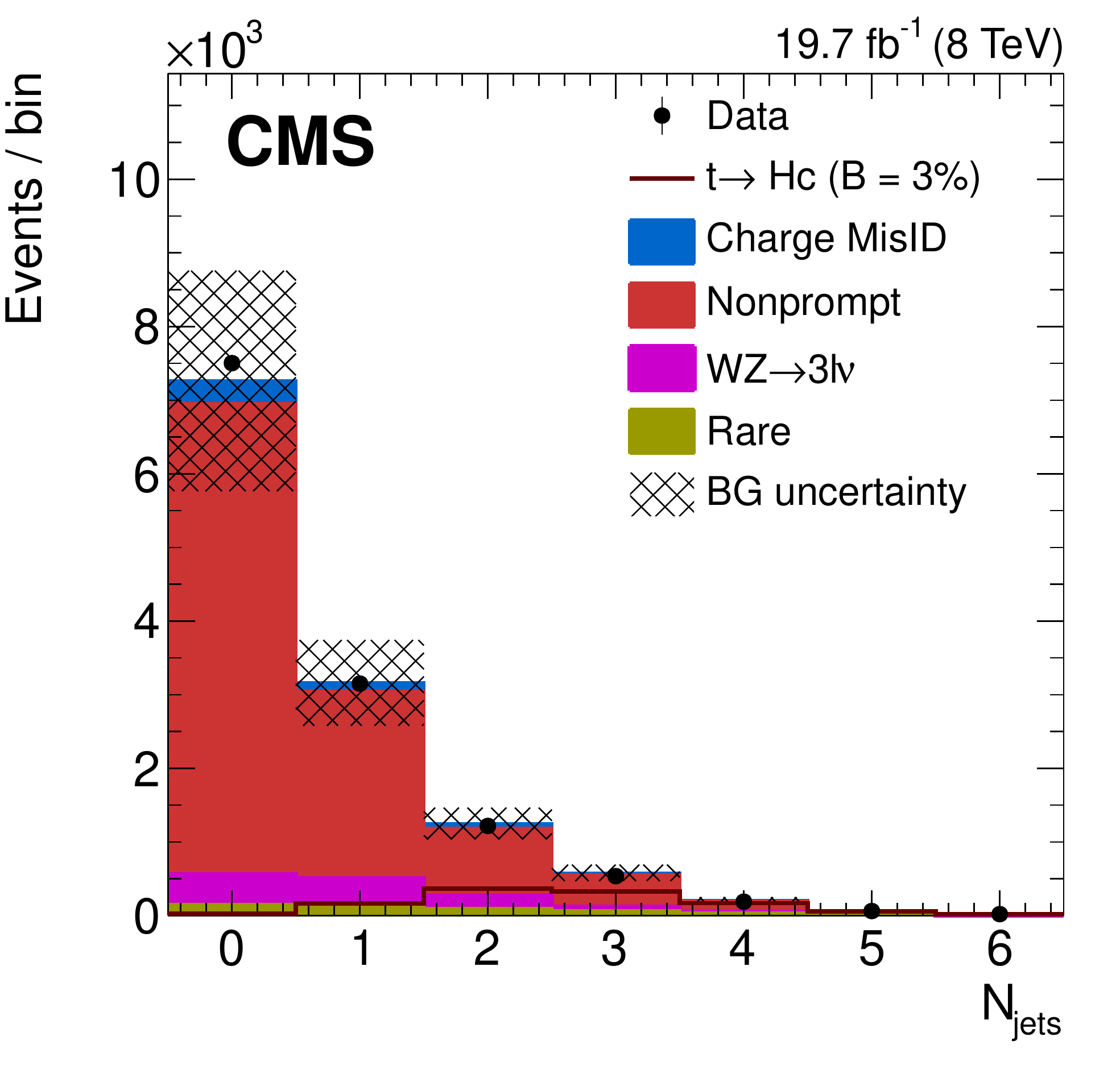}
    \caption{
    Jet multiplicity in the samples featuring three identified leptons
    (left) and two SS leptons (right) after rejecting events with \Z
    bosons.  The data are represented by the points with vertical bars,
    and the unfilled histogram shows the expected signal. A value of
    $\mathcal{B}({\rm t\to Hc}) = 3\%$ is used for the sake of improved
    visualization. The dominant backgrounds are represented with filled
    histograms and the background (BG) uncertainty is shown as shaded
    bands.
    }\label{fig:2}
    \end{center}
\end{figure}

\begin{figure}[h]
    \begin{center}
    \includegraphics[width=0.45\textwidth]{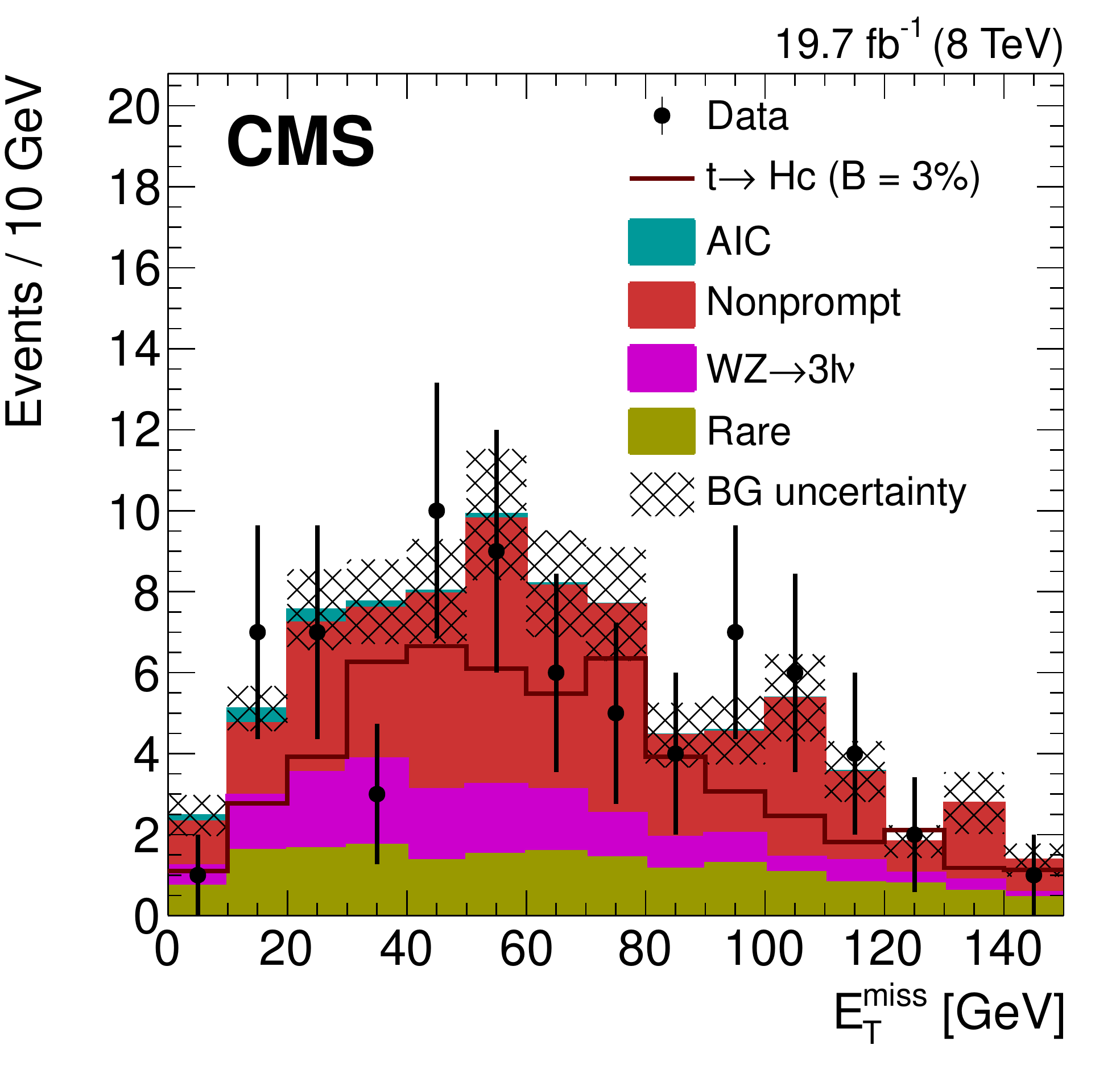}
    \includegraphics[width=0.45\textwidth]{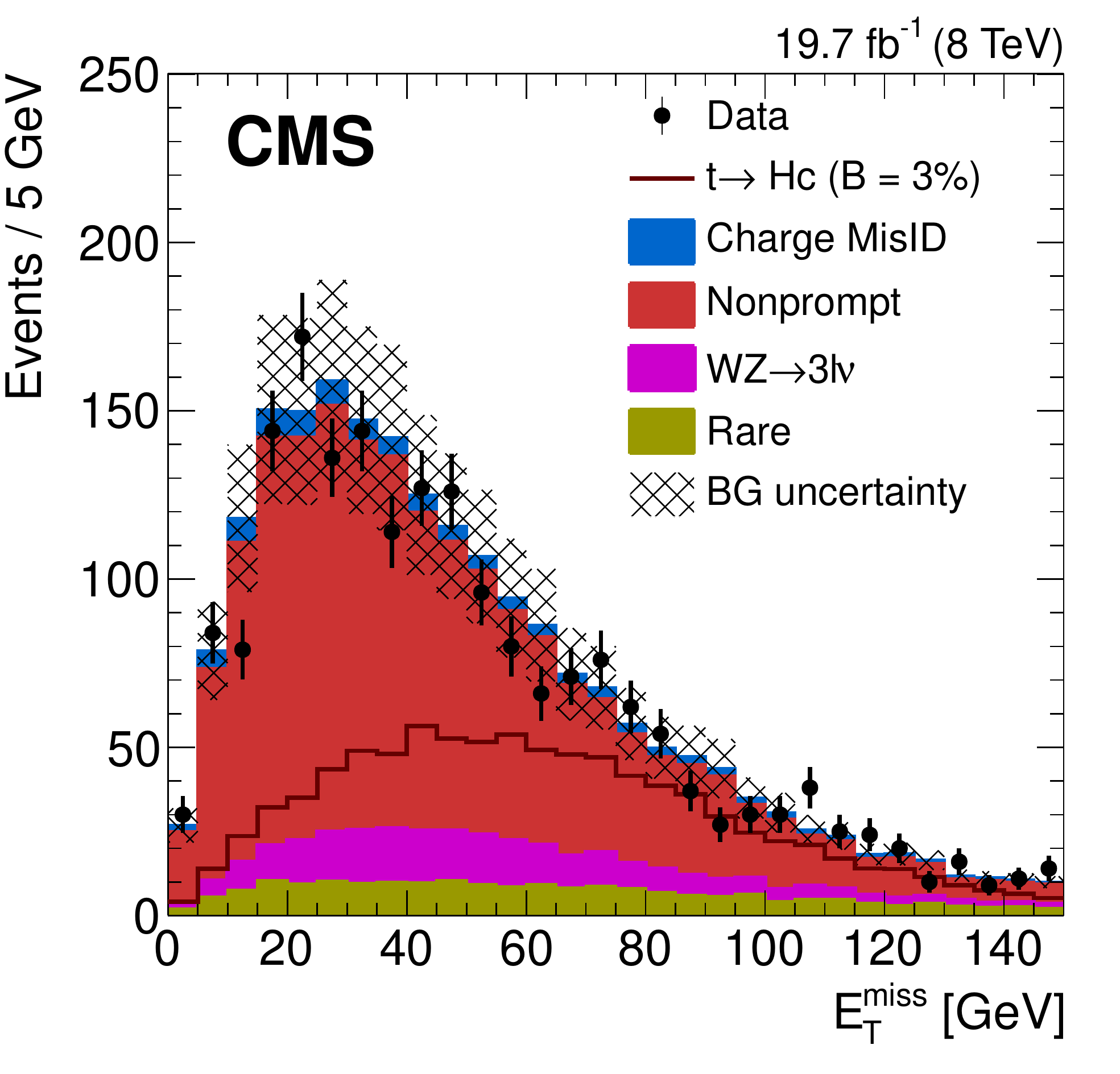}\\
    \includegraphics[width=0.45\textwidth]{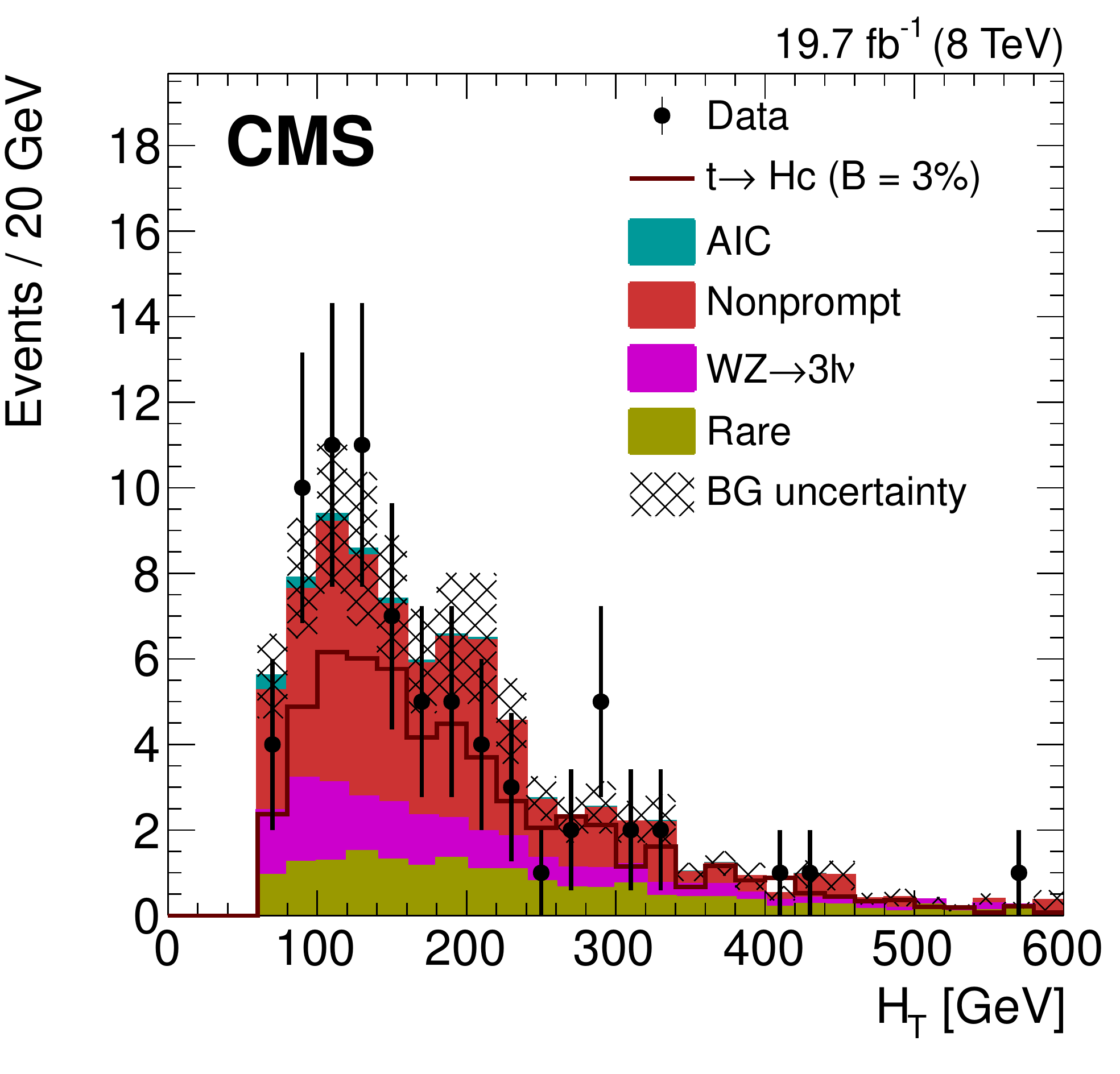}
    \includegraphics[width=0.45\textwidth]{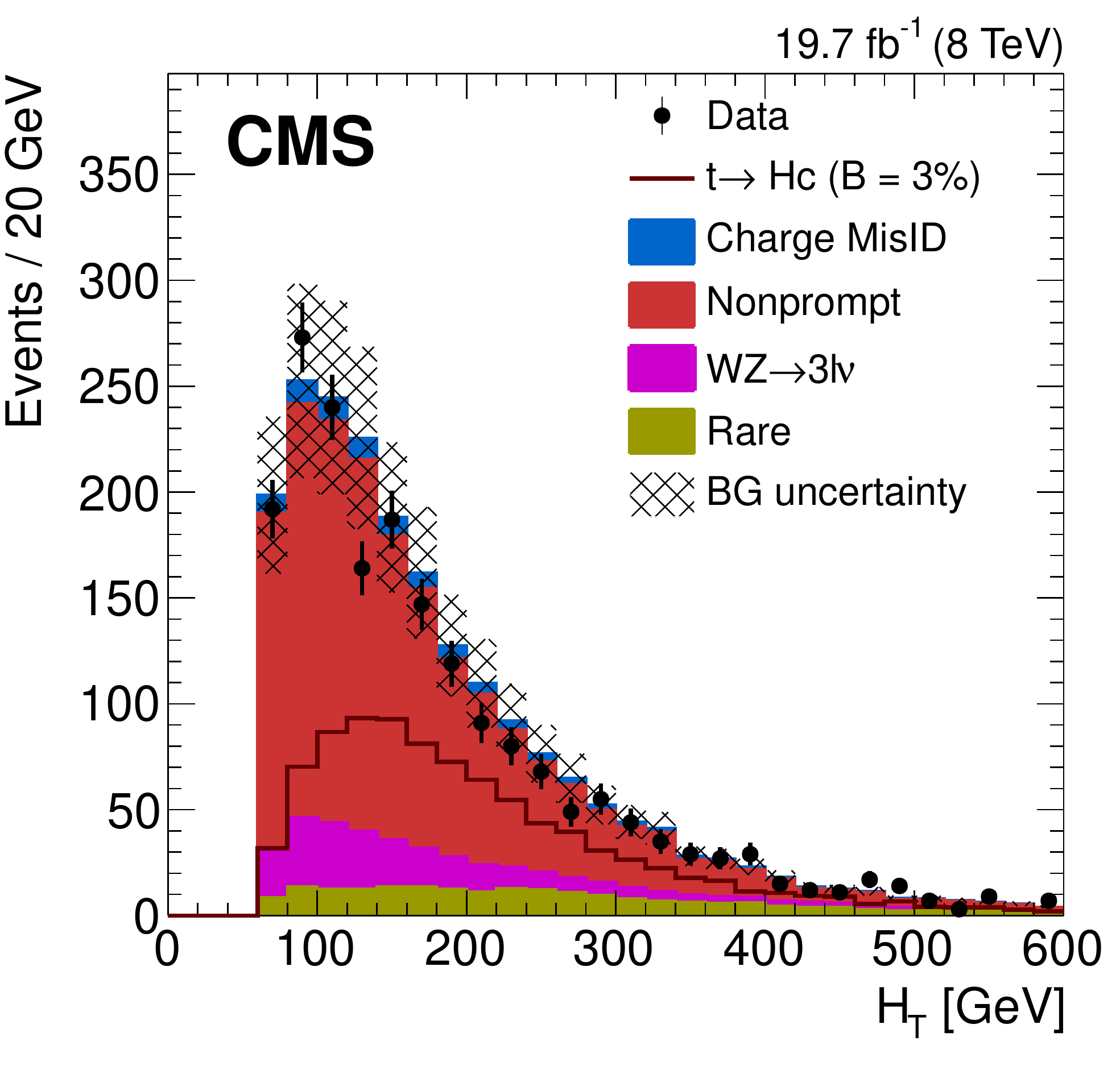}
    \caption{
    The \MET (top) and \HT (bottom) distributions in the trilepton (left)
    and SS dilepton (right) channels in data (points
    with bars) and predicted by the SM background simulations (filled
    histograms) after rejecting events containing \Z bosons, requiring at
    least two jets, and the event selection described in
    Section~\ref{sec:preselection}. The overall background uncertainty
    is shown in shaded black. The expected signal assuming a
    $\mathcal{B}({\rm t\to Hc})$ of 3\% is shown by the unfilled
    histogram.
    }\label{fig:3}
    \end{center}
\end{figure}

The jet multiplicity after rejecting events containing a \Z boson is
shown in Figure~\ref{fig:2}. To improve the sensitivity of the search, we
require at least two jets in the final state.
Figure~\ref{fig:3} shows the \MET and \HT distributions for trilepton
and SS dilepton events after applying the Z veto and jet requirement.  A
candidate event in the trilepton channel has no additional requirements
on \MET or \HT.  The SS events are required to pass an \MET-dependent
\HT requirement (shown in Table~\ref{table:MetHT}) and have \MET greater
than 30\GeV.  The \MET and \HT requirements are obtained by maximizing
the estimated signal significance, defined as the number of signal
events over the square root of the number of background
events.

The main sources of background can be divided into two categories
according to the origin of the identified leptons and the \MET.  These
include (1) \textit{irreducible background processes}: events with leptons
originating from the decay of SM bosons and having large \MET arising
from neutrinos; (2) \textit{reducible background processes:} events with
misidentified leptons produced either by nonprompt leptons from hadron
decays (\eg, semileptonic decays of B mesons), by misidentified hadrons,
or by mismeasurement of the lepton charge.

\begin{table}[ht]
    \begin{center}
    \topcaption{ Two-dimensional selection
    requirements on \MET and \HT applied in the SS dilepton channel. An
    event is selected if it satisfies one of the three listed sets.  }
    \label{table:MetHT}
    \renewcommand{\arraystretch}{1.1}
    \begin{tabular}{ c | c  c  c }
    Selection set      & 1 & 2& 3 \\ \hline \MET & $\phantom{1}{<}70\GeV$ & $70{-}90\GeV$ &
    ${>}90\GeV$ \\ \HT  & ${>}140\GeV$ & ${>}100\GeV$ & ${>}60\GeV$ \\
    \end{tabular}
    \end{center}
\end{table}

Given that at least two isolated leptons and two jets are required in
the final state, the main sources of irreducible backgrounds are
$\ttbar$ associated with vector boson production, \PW\Z$\to
3\ell\nu$, $\PZ\PZ\to 4\ell$, $\Z\to 4\ell$, and, to a
lesser extent, triboson and $\PW^{\pm}\PW^{\pm}$ production.  The
contribution from all of these processes except $\Z\to 4\ell$
production are estimated from simulated samples.  The WZ cross section
used in the simulation is cross-checked against a control sample from
data that is enriched in WZ events by requiring that there be three
leptons, with two of them forming a dilepton pair whose invariant mass
is consistent with a Z boson. No correction to the WZ normalization is
needed.  This sample is also used to assess the systematic uncertainty
in the simulation of the background.

For the presentation of the results, several of the backgrounds are
grouped into a single category referred to as the rare backgrounds.  The
rare background contribution is estimated mainly from simulation (see
the following paragraph), and the processes include $\PZ\PZ\to
4\ell$, $\ttbar$+Z, $\ttbar$+W, triboson, $\PW^{\pm}\PW^{\pm}$, and
$\ttbar$+$\PH$.  The $\PW\Z\to 3\ell\nu$ background contribution is
presented separately.

The residual contribution in the trilepton channel from asymmetric
internal conversions (AIC) arising from Drell--Yan events is estimated
using a data-driven technique~\cite{asy-conv} that uses $\Z\to
\ell^{+}\ell^{-}+\gamma$ events in data to model $\Z\to
\ell^{+}\ell^{-}+\text{e}/\mu$ events.  This is because the process that
gives rise to the two final states is the same (final-state radiation in
Drell--Yan events), and the third lepton that is detected in the AIC
event carries most of the photon momentum.  The
$\ell^{+}\ell^{-}+\gamma$ events are scaled based on photon
\pt-dependent weights coming from a control sample defined as having a
three-body invariant mass within 15\GeV of the Z boson mass.  The
average conversion probabilities for photons in dimuon and dielectron
events are $(0.57 \pm 0.07)\%$ and $(0.7 \pm 0.1)\%$, respectively.

There are two major types of reducible backgrounds coming from \bbbar,
Drell--Yan, W+jets, and $\ttbar$ processes.  One source comes from
events with either nonprompt leptons produced during the hadronization
process of the outgoing quarks (\eg, semileptonic decays of B mesons) or
hadrons misidentified as prompt leptons.  The other source originates
from the charge misidentification of a lepton in the more frequent
production of opposite-sign dileptons. This background mostly
contaminates the SS dielectron final states.  Data-driven methods are
used to estimate these two types of reducible backgrounds.

Mismeasuring the charge of a lepton can be a significant source of
background in SS dilepton final states when there are one or more
electrons.  Even though the probability for mismeasuring the charge of
an electron is relatively low (${\approx}0.1\%$), the production rate of
opposite-sign dileptons is very high in comparison to processes that
result in genuine SS dileptons. The probability of mismeasuring the
charge of a muon is negligible (${<}10^{-6}$) and is therefore not
considered here.  In order to estimate the probability of misidentifying
the charge of an electron from data, a control sample is selected
consisting of events containing a dielectron pair with an invariant mass
within 15\GeV of the \Z boson mass.  The rate of charge
misidentification is then determined from the ratio of the number of SS
events to opposite-sign events as a function of \pt and $\eta$.  The
measured charge misidentification for electrons with $\abs{\eta} <
1.48$ is less than 0.2\% for $\pt < 100\GeV$, while for
$\abs{\eta} > 1.48$ it is 0.1\% at 10\GeV and increases with
\pt to 2.5\% at 125\GeV.  These measurements are in agreement with those
obtained from simulated Drell--Yan events.

Two control samples are used to estimate the misidentification rate of
prompt
leptons~\cite{Chatrchyan:2011wba,Chatrchyan:2012ira,Chatrchyan:2012ty}:
one region is enriched in \bbbar events; the other is enriched
in Z + jet production.  Both samples are used to estimate the
probability of misidentifying nonprompt electrons and muons as a
function of $\pt$ and $\eta$.  The measured misidentification rate for
electrons ranges from 2\% to 8\% and for muons ranges from 1\% to 6\%.
Simulated events are used to correct for the contamination arising from
prompt leptons in the nonprompt misidentification rate measurement
(e.g., WZ production in the Z+jet control region).  The rates are then
applied to events where one or more of the lepton candidates fail the
tight lepton identification requirements.  The differences between the
nonprompt misidentification rates in the two measurement regions and the
signal region are then used to estimate the systematic uncertainty of
this background.  To further assess the systematic uncertainty, the
misidentification rates are also measured in simulated events that
reproduce the background composition of events in the signal region and
compared to the rates measured from data.

The predicted numbers of background and signal events for the trilepton
and SS dileptons are given in Table~\ref{table:yieldsa}. The backgrounds
are separated into nonprompt lepton, charge misidentification,
$\PW\PZ\to 3\ell\nu$, and the rare backgrounds.  The predicted
number of signal events assumes $\mathcal{B}({\rm t\to Hq}) =
1\%$.  The total number of observed events, also given in
Table~\ref{table:yieldsa}, is consistent with the predicted number of
background events.

\begin{table}[t]
    \begin{center}
    \topcaption{
    The predicted and observed inclusive event yields after the full
    event selection for the trilepton and SS dilepton
    categories assuming $\mathcal{B}({\rm t\to Hq}) = 1\%$.
    The quoted uncertainties include both statistical and systematic
    uncertainties added in quadrature.  The total number of observed
    events is given in the last row.
    }\label{table:yieldsa}
    \begin{tabular}{ l | c c | c c }
    Process                  & \multicolumn{2}{c|}{Trilepton} 	  & \multicolumn{2}{c}{SS dilepton} \\
    \hline
    Nonprompt                & \multicolumn{2}{c|}{$49.4 \pm 9.0$} & \multicolumn{2}{c}{$409 \pm 72$} \\
    Charge misidentification & \multicolumn{2}{c|}{---}             & \multicolumn{2}{c}{$32.1 \pm 6.4$} \\
    $\mathrm{WZ}\to3\ell\nu$  & \multicolumn{2}{c|}{$15.8 \pm 1.1$} & \multicolumn{2}{c}{$83.9 \pm 5.4$} \\
    Rare backgrounds         & \multicolumn{2}{c|}{$19.6 \pm 1.4$} & \multicolumn{2}{c}{$128.1 \pm 6.4\x$} \\
    \hline
    Total background         & \multicolumn{2}{c|}{$86.2 \pm 9.3$} & \multicolumn{2}{c}{$654 \pm 73$} \\
    \hline
    Signal                   & t$\to$Hu       & t$\to$Hc       & t$\to$Hu       & t$\to$Hc \\
    \hline
    H$ \to$ WW       & 12.4 $\pm$ 1.4 & 14.4 $\pm$ 1.1 & 135 $\pm$ 12   & 130.3 $\pm$ 8.1\x  \\
    H$ \to \tau\tau$ & \x4.1 $\pm$ 0.4  & \x4.4 $\pm$ 0.3  & 36.4 $\pm$ 3.2 & 35.3 $\pm$ 2.2  \\
    H$ \to$ ZZ       & \x0.4 $\pm$ 0.1  & \x0.4 $\pm$ 0.1  & \x1.6 $\pm$ 0.1  & \x1.4 $\pm$ 0.1  \\
    \hline
    Total signal             & 16.9 $\pm$ 1.5 & 19.2 $\pm$ 1.1 & 173 $\pm$ 13   & 167.0 $\pm$ 8.4\x  \\
    \hline\hline
    Observed                 & \multicolumn{2}{c|}{79}  & \multicolumn{2}{c}{631}  \\
    \end{tabular}
    \end{center}
\end{table}

\subsection{Diphoton channel}

The diphoton analysis is performed using both leptonic and hadronic W
boson decays: $\ttbar \to \PH\cPq + \PW\cPqb \to \gamma\gamma\cPq +
\ell\nu\cPqb$, and $\ttbar \to \PH\cPq + \PW\cPqb \to \gamma\gamma\cPq +
\cPq\cPq\cPqb$.  The mass of the diphoton system $m_{\gamma\gamma}$ is
the primary variable used to search for the Higgs boson decay. The
contribution of the nonresonant backgrounds is estimated by fitting the
$m_{\gamma\gamma}$ distribution from data in the mass range $100 <
m_{\gamma\gamma}<  180\,\GeV$, whereas the contribution of resonant
backgrounds is taken from the simulation.

The two highest-\pt photons must have $\pt > m_{\gamma\gamma}/3$
and $\pt > m_{\gamma\gamma}/4$, respectively.  The use of $\pt$
thresholds scaled by $m_{\gamma\gamma}$ prevents a distortion of the low
end of the $m_{\gamma\gamma}$ spectrum that would result from a fixed
threshold~\cite{Chatrchyan:2012twa}.  In the rare case of multiple
diphoton candidates in an event, the one with the highest $\pt$ sum is
selected.

The hadronic analysis uses events with at least four jets and exactly
one b jet.  The b jet and the three jets with the highest $\pt$ are used
to reconstruct the invariant mass of the two top quarks, $m_{{\rm
j}\gamma\gamma}$ and  $m_{{\rm bjj}}$.  There are three possible
$(m_{{\rm j}\gamma\gamma}, m_{{\rm bjj}})$ pairs per event.  The
combination of jets with the minimum value of $\left| m_{{\rm
j}\gamma\gamma}/ m_{{\rm bjj}} -1 \right| + \left| m_{{\rm bjj}}/m_{{\rm
j}\gamma\gamma} -1\right|$ is selected.  The allowed ranges for $m_{{\rm
j}\gamma\gamma}$, $m_{{\rm bjj}}$, and the W boson mass $m_{\PW}$
associated with $m_{{\rm bjj}}$ are obtained by maximizing the signal
significance $S/\sqrt{B}$ in the simulation, where $S$ is the number of
signal events and $B$ is number of the background events. The background
events are assumed to come from $\gamma\gamma$+jets and are taken from
simulation.  The highest signal significance is found to be 16\%
obtained for $142 \le m_{{\rm bjj}} \le 222\GeV$, $158 \le m_{{\rm
j}\gamma\gamma} \le 202\GeV$, and $44\le m_{\rm W} \le 140\GeV$.

The leptonic analysis uses events with at least three jets, exactly one
b jet, and at least one lepton.  The reconstructed top mass $m_{{\rm
b}\nu\ell}$ is found from the b jet, the lepton, and \MET.  The
longitudinal momentum of the neutrino is estimated by using the W boson
mass as a constraint, which leads to a quadratic equation. If the
equation has a complex solution, the real part of the solution is used.
If the equation has two real solutions, the one with the smaller value
of $\left| m_{{\rm j}\gamma\gamma}/ m_{{\rm b}\ell\nu} -1 \right|
+\left| m_{{\rm b}\ell\nu}/m_{{\rm j}\gamma\gamma}  -1\right|$ is
chosen.  The mass windows for $m_{\rm bjj}$, $m_{\rj\gamma\gamma}$, and
$m_{\PW}$ are the same as in the hadronic channel.

The signal region is defined using the experimental width of the Higgs
boson, 1.4\GeV, around the nominal mass peak position. As in the
analysis of the inclusive SM Higgs boson decaying into
diphotons~\cite{Chatrchyan:2012twa}, the signal shape of the diphoton
invariant mass distribution is described by the sum of three Gaussian
functions.  Although the contribution from the SM Higgs boson
background, dominated by the $\ttbar$H process, is relatively small in
comparison to the contribution of the nonresonant diphoton background,
the resonant diphoton background cannot be ignored because it has a very
similar $m_{\gamma\gamma}$ distribution as the signal.

To determine the shape of the nonresonant diphoton background,
a function consisting of a test model and the resonant diphoton
background is fitted to the data under the background-only hypothesis. The
model of the resonant diphoton background is the same as the signal
function. The background function is used to generate 1000 pseudo-experiment
samples that are fitted with the background plus signal probability density
function.

\begin{figure}[h]
    \begin{center}
    \includegraphics[width=0.45\textwidth]{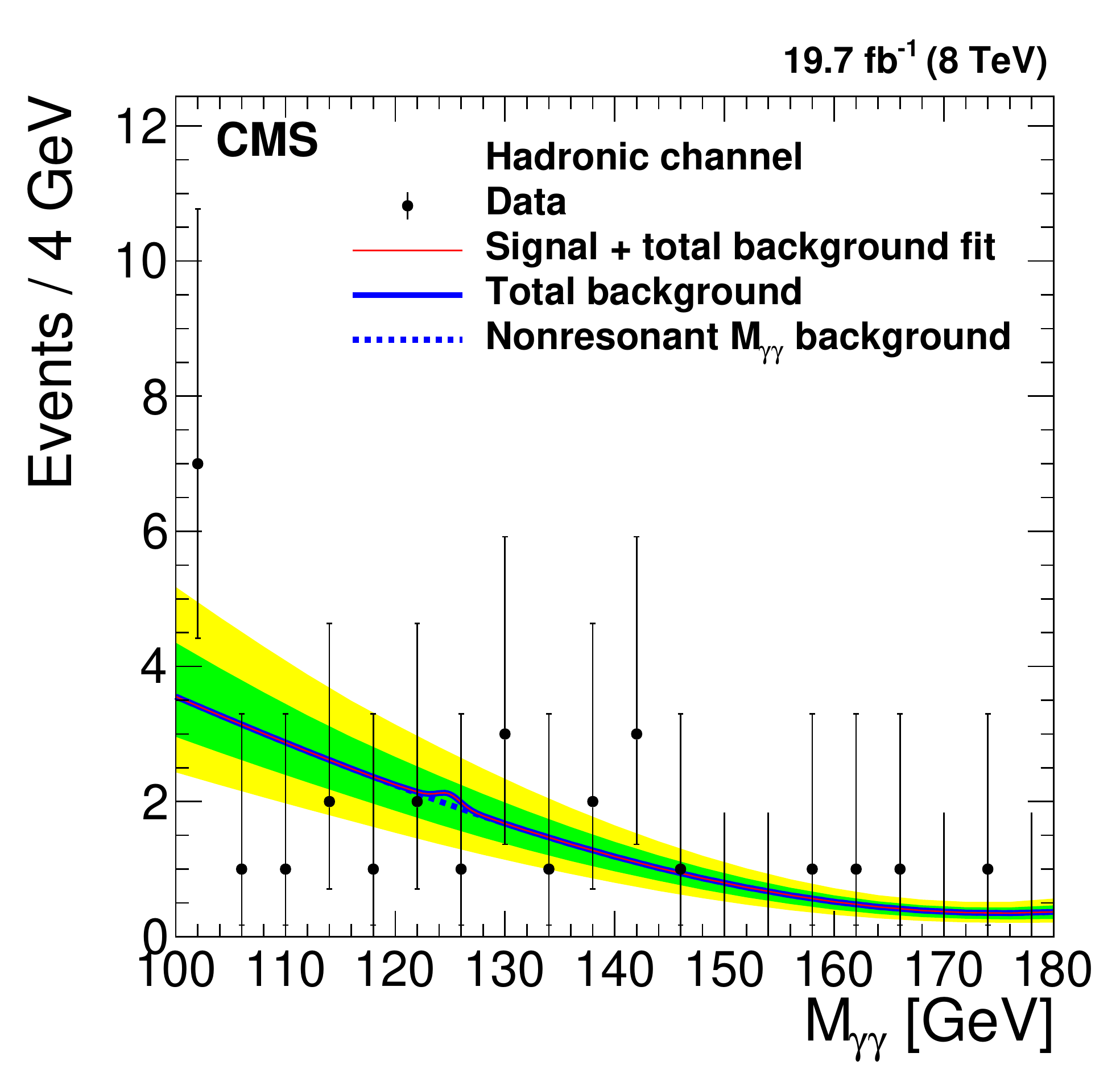}
    \includegraphics[width=0.45\textwidth]{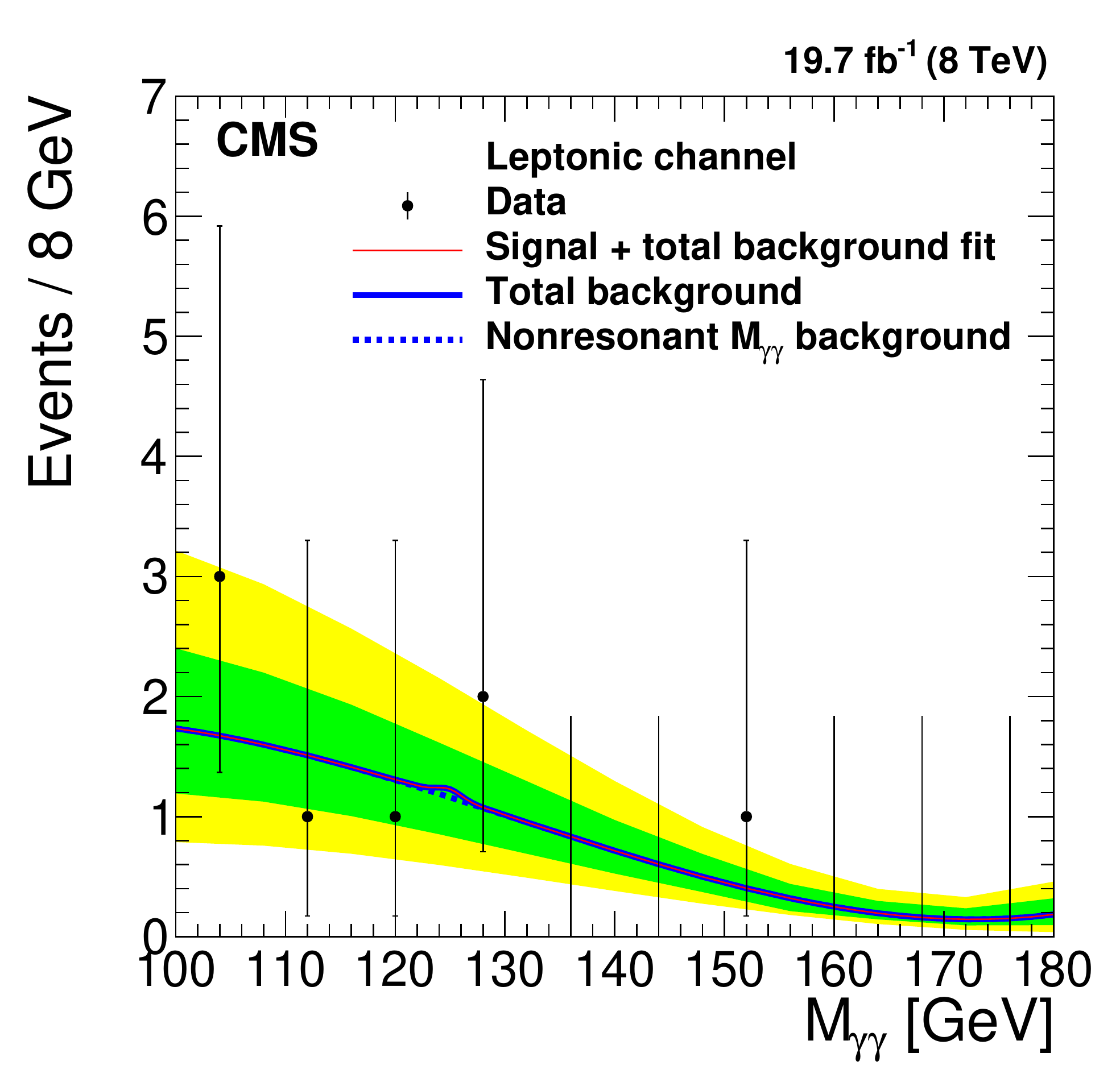}
    \caption{\label{fig:gghl}
    The $m_{\gamma\gamma}$ distribution and the fit result of the
    hadronic (left) and leptonic (right) channels. The dashed
    line represents the component of the nonresonant diphoton
    background, while the solid line represents the total background
    plus signal. The shaded bands represent one and two
    standard deviation uncertainties of the fit.}
    \end{center}
\end{figure}

A pull is then defined as $(N_{\rm fit}-N_{\rm gen})/\sigma_{N_{\rm
fit}}$, where $N_{\rm fit}$ is the fitted number of signal events in the
pseudo-experiments, $N_{\rm gen}$ is the number of generated signal
events, and $\sigma_{N_{\rm fit}}$ is the corresponding uncertainty. In
the case under consideration, $N_{\rm gen} = 0$.  The procedure is
verified by injecting signal in the pseudo-experiments.  Several models
are tried, and the chosen function for nonresonant diphoton background
is the one whose bias (offset of the pull distribution) is less than
0.15 and with the minimum number of degrees of freedom for the entire
set of tested models.  A third-order Bernstein polynomial is selected as
the functional form of the background for both the hadronic and leptonic
channels.  After determining the function to describe the nonresonant
diphoton background, a function given by the sum of probability density
functions of the resonant and nonresonant diphoton backgrounds and signal
is fitted to the data. The normalization of the resonant diphoton
background is allowed to vary within its uncertainties, while the
normalization of the nonresonant component is unconstrained.
Table~\ref{tab:gg-s-bg} gives a summary of the observed and expected
event yields for the two diphoton channels and Figure~\ref{fig:gghl}
shows the fit result overlaid with the data.

\begin{table}[h]
    \topcaption{
    Observed event yield and the expected numbers of background and
    signal events for the diphoton selection in the hadronic and
    leptonic channels in the $100<m_{\gamma\gamma} < 180\GeV$ mass
    range.  The signal yields assume $\mathcal{B}(\cPqt\to\PH\cPq)
    = 1\%$.  The uncertainties are statistical only.
    \label{tab:gg-s-bg}}
    \begin{center}
    \renewcommand{\arraystretch}{1.1}
    \begin{tabular}{l | c c}
    Process                 & Hadronic channel & Leptonic channel\\
    \hline
    Nonresonant background  & 28.9 $\pm$ 5.4\x   & 8.0 $\pm$ 2.8 \\
    Resonant background     & 0.15 $\pm$ 0.02  & 0.04 $\pm$ 0.01 \\
    \hline
    $\cPqt\to\PH\cPqc$      & 6.26 $\pm$ 0.07  & 1.91 $\pm$ 0.04 \\
    $\cPqt\to\PH\cPqu$      & 7.09 $\pm$ 0.08  & 2.02 $\pm$ 0.04 \\
    \hline\hline
    Observed & 29 & 8\\
    \end{tabular}
    \end{center}
\end{table}

\subsection{b jet + lepton channel}

\begin{figure}[ht]
    \begin{center}
    \vspace{1cm}
    \includegraphics[width=0.45\textwidth]{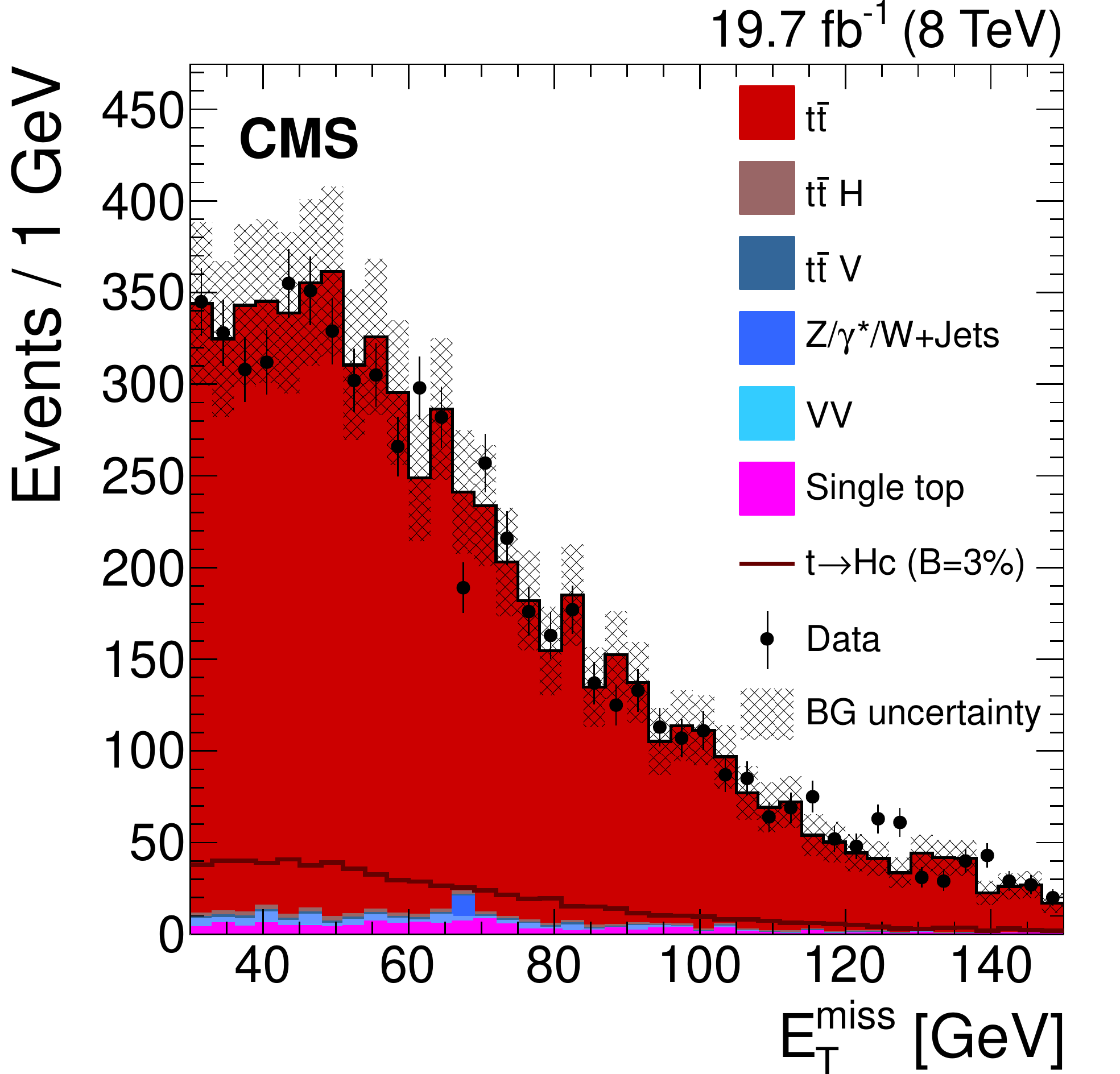}
    \includegraphics[width=0.45\textwidth]{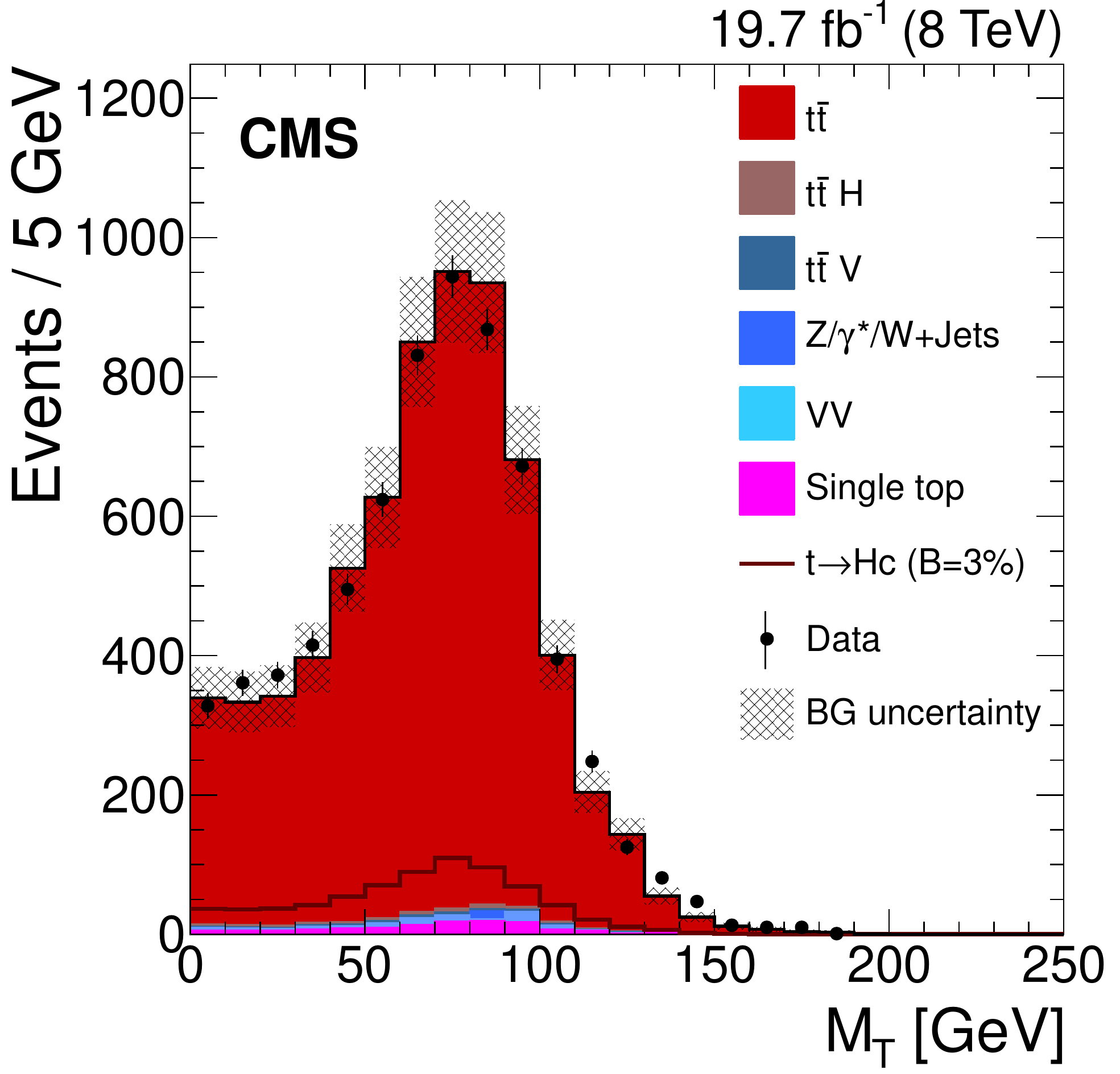}
    \caption{
    Comparison between data and simulated events after the basic
    selection for b jet + lepton events has been applied: the \MET
    distribution (left) and the reconstructed transverse mass of the W
    boson candidate (right). A value of $\mathcal{B}({\rm t\to Hc}) =
    3\%$ is used for the sake of improved visualization.
    }\label{fig:1c}
    \end{center}
\end{figure}

The basic event selection requirements for the b jet + lepton channel
are a single-lepton trigger, one isolated lepton, a minimum \MET of
30\GeV, and at least four jets, with at least three of them tagged as b
jets.  The background is dominated by $\ttbar \to
\bbbar\PW^{+}\PW^{-}$ production.  Figure~\ref{fig:1c} shows the
distributions of \MET and the W boson transverse mass ($M_{\mathrm{T}}$) for data
and simulation after the basic event selection criteria are applied.
The transverse mass is defined as

\begin{equation*}
    M_{\mathrm{T}} = \sqrt{2\pt^{\ell}\MET[1 - \cos(\Delta\phi(\ell, \nu))]},
\end{equation*}

where $\pt^{\ell}$ is the \pt of the lepton, \MET is used in place of
the \pt of the neutrino, and $\Delta\phi(\ell, \nu))$ is the azimuthal
angular difference between the directions of the lepton and neutrino.

For both top quark decays $\cPqt \to \PH\cPq \to \bbbar \rj$ and $\cPqt
\to \PW\cPqb \to \cPqb\ell\nu$, a full reconstruction of the top
quark invariant mass $m_{\PH\cPq}$ or $m_{\PW\cPqb}$ is possible. However,
combinatorial background arises since there is no unambiguous way to
match multiple light-quark and b quark jets with the final-state
quarks.  Therefore, all possible combinations are examined and a
multivariate analysis (MVA) technique~\cite{Hocker:2007ht} is used to
select the best candidate for each event. Several variables based on
event kinematics and event topology are examined.  Considering their
signal-to-background separation power, the following variables are used
to form a boosted decision tree (BDT)~classifier~\cite{Hocker:2007ht}:

\begin{itemize}
    \item the invariant masses $m_{\PH\cPq}$ and $m_{\PH\cPqb}$ of the reconstructed top quarks,
    \item the energy of the u or c jet from the
    $\cPqt\to\cPq\PH$ in the rest frame of its parent top quark,
    \item the azimuthal angle between the reconstructed top quarks
    directions,
    \item the azimuthal angle between the reconstructed W boson and
    the associated b jet directions,
    \item the azimuthal angle between the Higgs boson and the
    associated jet directions,
    \item the azimuthal angle between the directions of the b jets resulting from the
    Higgs boson decay.
\end{itemize}

The BDT classifier is trained with the correct and wrong combinations of
simulated FCNC events determined from the generator-level parton
matching. Because only event kinematics and topological variables are
used, the Hu and Hc channels share the same BDT classifier.  The
jet-parton assignment in each event is determined by choosing the
combination with the largest BDT classifier score, resulting in the correct
assignment in 54\% of events, as determined from simulation.
The signal is determined using a template fit of the output of an
artificial neural network (ANN)~\cite{Hocker:2007ht}.  The ANN takes its
inputs from the invariant mass of the reconstructed Higgs boson
candidate and the CSV discriminator variables of the three b jets from
the hadronic top quark and Higgs boson daughters.  The training of the
ANN is done separately for the $\cPqt\to\PH\cPqu$ and
$\cPqt\to\PH\cPqc$ channels. A control sample dominated by
$\ttbar$ is selected to validate the simulation used in the training.
The sample is constructed by requiring one lepton and four jets, of
which exactly two are b jets.

\begin{figure}[h]
    \begin{center}
    \includegraphics[width=0.45\textwidth]{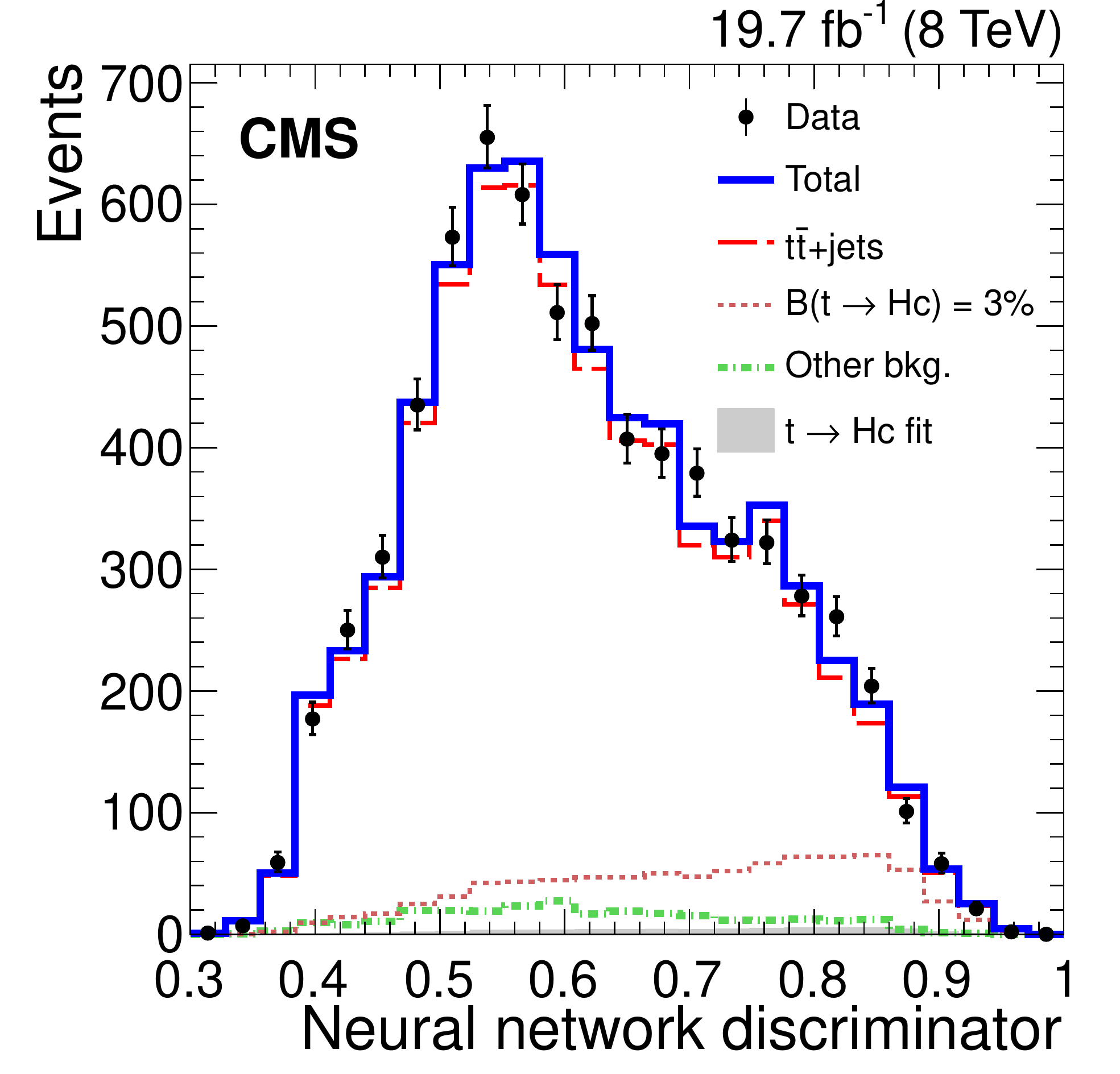}
    \includegraphics[width=0.45\textwidth]{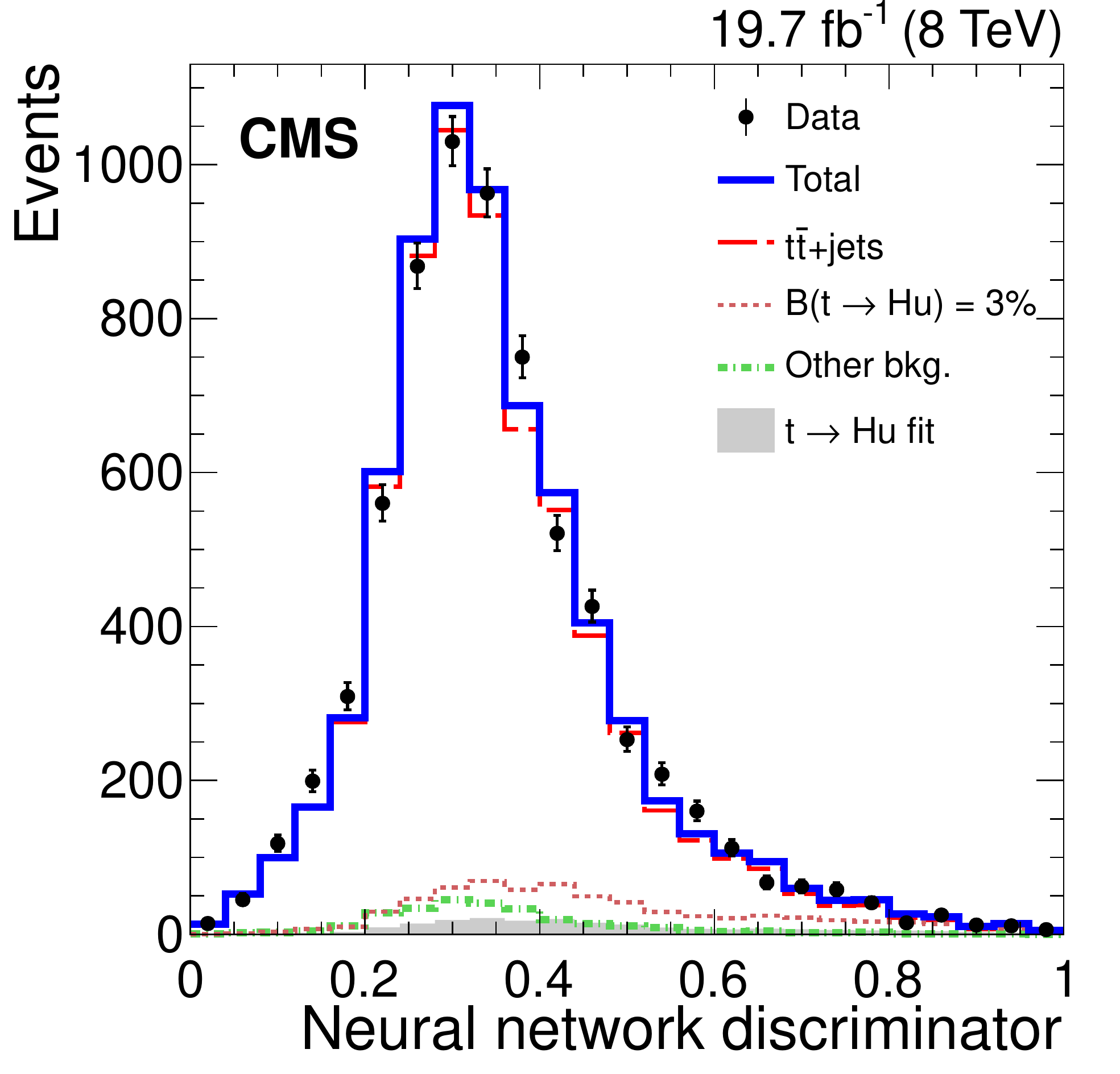}
    \caption{
    The output distributions from the ANN discriminator for data
    (points) and simulated background (lines) where the ANN was
    trained to discriminate the backgrounds from either
    $\cPqt\to\PH\cPqc$ (left) or $\cPqt\to\PH\cPqu$ (right) decays.
    The solid line shows the result of the fit of the signal and
    background templates to data.  The dotted line gives the
    predicted signal distribution from simulation for
    $\mathcal{B}(\cPqt\to\PH\cPqc) = 3\%$ and the filled histogram
    shows the proportion of signal estimated from the fit.
    \label{plot:bb-fit}}
    \end{center}
\end{figure}

Figure~\ref{plot:bb-fit} show the results of the fit performed with the
6840 observed events.  The observed number of events and the expected
yields of the signal and the main backgrounds estimated from simulation
are shown in Table~\ref{table:multijet_sim_results}.  The estimated
background and signal based on the fit of the ANN discriminator output
is shown in Table~\ref{table:multijet_fit_results}.  The number of
signal and background events from the fit result for the Hc channel are
$74 \pm 109\stat\pm~24\syst$ and $6770 \pm 130\stat\pm~950\syst$, respectively. The corresponding yields for the Hu channel are
$197 \pm 87\stat\pm~59\syst$ and $6640 \pm 120\stat\pm~800\syst$, respectively.

\begin{table}[h]
    \topcaption{The expected number of background and signal events for the
    b jet + lepton selection from simulation.  The signal yields from
    the simulation of the signal assume $\mathcal{B}(\cPqt\to\PH\cPq) =
    1\%$.  Uncertainties combine both statistical and
    systematic components in quadrature.
    \label{table:multijet_sim_results}}
    \centering
    \begin{tabular}{l | c}
    Process & Predicted number of events \\
    \hline
    \ttbar             & $7100 \pm 1500$ \\
    \ttbar\PH          & $55 \pm 11$     \\
    \PW\bbbar          & $71 \pm 14$     \\
    \hline
    Total background   & $7226 \pm 1500$ \\
    \hline
    $\cPqt\to\PH\cPqc$ & $272 \pm 90\x$    \\
    $\cPqt\to\PH\cPqu$ & $215 \pm 65\x$    \\
    \hline\hline
    Observed & 6840 \\
    \end{tabular}
\end{table}

\begin{table}[h]
    \topcaption{The measured number of background and signal events for the
    b jet + lepton selection from fitting the ANN output
    trained on $\cPqt\to\PH\cPqc$ and $\cPqt\to\PH\cPqu$ final states.
    Uncertainties are statistical and systematic values,
    respectively.  The observed number of events is shown in the last
    row.
    \label{table:multijet_fit_results}}
    \centering
    \begin{tabular}{l | c c}
    Process    & $\cPqt\to\PH\cPqc$     & $\cPqt\to\PH\cPqu$ \\
    \hline
    Background & $6770 \pm 130 \pm 950$ & $6440 \pm 120 \pm 800$ \\
    Signal     & $\x74 \pm 109 \pm 24$    & $\x197 \pm \x87 \pm 59\x$ \\
    \hline\hline
    Observed & \multicolumn{2}{c}{6840} \\
    \end{tabular}
\end{table}

\section{Systematic uncertainties}
\label{sec:sys}

In the fit to the data, systematic uncertainties are treated as nuisance
parameters.  Each of them is assigned  a log-normal or Gaussian pdf,
which is included into the likelihood in a frequentist manner by
interpreting it as signal arising from pseudo-measurement distributions.
Nuisance parameters can affect either the signal yield, the shape of
kinematic variable distributions, or both.  If a specific source of
uncertainty is not included for a given channel,  it indicates that the
uncertainty is either not applicable to that channel or is found to have
negligible impact on the result.

The sources of uncertainties common to all analysis channels are: the
uncertainty in the total integrated luminosity
(2.6\%)~\cite{CMS-PAS-LUM-13-001}; the effects of the event pileup
modeling for the signal samples (0.2--3\%), which is particularly
important for the b jet + lepton channel; the uncertainty in the Higgs
boson branching fractions (5\%)~\cite{Denner:2011mq}; the uncertainty in the \ttbar cross
section (7.5\%)~\cite{Khachatryan:2014loa}; the uncertainty in the jet
energy scale (1--15\%)~\cite{ref:jetscale} and resolution (0.4--8\%),
where the larger uncertainty is for the b jet + lepton selection; the
uncertainty in the PDF used in the event generators
($<9\%$)~\cite{Bourilkov:2006cj}; the assumed top quark \pt distribution
(1--4\%)~\cite{CMS-PAS-TOP-12-027}; the \MET resolution
(0.2--4\%)~\cite{ref:jetscale}; the uncertainty in the trigger
efficiency (${<}2\%$); and the corrections applied to the simulation to
account for the differences in lepton identification and isolation
efficiencies in data and simulation (0.01--6\%), where the larger
uncertainty is for the selection of events with a three-electron final
state.

The uncertainties specific to the signal description and background
estimation for the multilepton analysis come from the 11--13\%
uncertainty in the $\ttbar$W and $\ttbar$Z theoretical cross
sections~\cite{Garzelli:2012bn}; the 15\% uncertainty in the WZ
normalization (determined from a control region); the uncertainty in the
lepton misidentification rate (40\% for electrons, 30\% for muons); and
the 20\% uncertainty in the electron charge mismeasurement probability.
The uncertainties specific to the signal description and background estimation
for the diphoton channels are the corrections applied to the
simulation to account for differences of the photon identification efficiency
in data and simulation (0.1--5\%); and the uncertainty in the jet and b jet
identification efficiency (2--3.5\%)~\cite{CSV}.  The resonant background from
the SM Higgs boson production has an uncertainty of 8.1\% from the PDF
uncertainty and 9.3\% from the QCD scale~\cite{lhcwg12}.

The uncertainties specific to the signal description and background
estimation for the b jet + lepton channel are dominated by the b jet
identification. The uncertainty in the b tagging correction has two
components: one is from the sample purity (4\%)~\cite{CSV} and the other
from the sample statistical uncertainty (24\%).  The uncertainty in
the $\ttbar$+jets cross section, determined using a leading-order event
generator, is 1\%. The uncertainty in the modeling of the heavy-flavor
daughters of the W decay in the \ttbar simulated sample is estimated to
be 3\%. Additional uncertainties arise from  the event generator
parameters such as the renormalization and factorization scales
(5\%)~\cite{PhysRevD.78.013004}, the parton-jet matching threshold (1--9\%), and
the top quark mass (4\%).

The uncertainties owing to the integrated luminosity, jet energy scale
and resolution, pileup, reconstruction of physics objects, signal PDFs,
and top quark related uncertainties are assumed to be fully correlated,
while all others are treated as uncorrelated.

The systematic uncertainties are summarized in Table~\ref{sys}.

\begin{table}[ht!]
    \topcaption{
    Systematic uncertainties for the $\rm t\bar{t} \to Hq  + Wb$~(q
    = u, c) channels in percent.  Ranges are quoted to indicate
    values that vary across the different analyses.
    } \label{sys}
    {
    \begin{center}
    \renewcommand{\arraystretch}{1.05}
    \resizebox{\textwidth}{!}{
    \begin{tabular}{l|ccccc}
    Channel  & SS dilepton & Trilepton & $\gamma\gamma$ hadronic& $\gamma\gamma$ leptonic & b jet + lepton \\
    \hline
    Integrated luminosity                   & 2.6        & 2.6        & 2.6 & 2.6 & 2.6    \\
    Pileup                                  & 1.0        & 1.0 &   0.3 & 0.8 & 0.2--3.0 \\
    Higgs boson branching fraction          & 5.0        & 5.0        & 5.0 & 5.0 & 5.0\\
    ${\rm t\bar{t}}$ cross section          & 7.5        & 7.5        & 7.5 & 7.5 & 7.5\\
    Jet energy scale                        & 0.5        & 0.6        & 1.2 & 1.0 & 5.2--15\phantom{.}\\
    Jet energy resolution                   & 0.8        & 2.2        & 2.7 & 0.4 & 2.2--7.8\\
    Signal PDF                              & 6.0        & 6.0        & 5.9 & 5.2 & ${<}1$--9.0\\
    Top quark $p_{\rm T}$ correction & ---         & ---         & 1.4 & 3.2 & 0.8--4.3\\
    \MET                                    & 4.0        & 4.0        & ---  & ---  & 0.2--2.5 \\
    Trigger efficiency                      & 1.0--2.0 & ---         & 1.0 & 1.0 & ${<}0.1$--0.4\x\\
    Identification and  isolation           &            &            &     &     & \\
    ~~~~- muon                              & 1.0--2.0 & 1.0--3.0 & ---  & 0.3 & 0.01--0.04 \\
    ~~~~- electron                          & 2.0--4.0 & 2.0--6.0 & ---  & 0.3 & ${<}0.1$--0.2\x \\
    \hline
    ${\rm t\bar{t}}$W normalization & 11.0 & 11.0 & --- & --- & --- \\
    ${\rm t\bar{t}}$Z normalization & 13.0 & 13.0 & --- & --- & ---\\
    WZ normalization                & 15.0 & 15.0 & --- & --- & ---\\
    Lepton misidentification        &      &      &    &    & \\
    ~~~~- electron                  & 40.0 & 40.0 & --- & --- & ---\\
    ~~~~- muon                      & 30.0 & 30.0 & --- & --- & ---\\
    Charge misidentification        & 20.0 & ---   & --- & --- & ---\\
    \hline
    Photon identification efficiency  & --- & --- & 5.2 & 5.2 & --- \\
    Corrections  per photon           &    &    &     &     & \\
    ~~~- energy scale                 & --- & --- & 0.1 & 0.1 & --- \\
    ~~~- energy resolution            & --- & --- & 0.1 & 0.1 & --- \\
    ~~~- material mismodeling         & --- & --- & 0.3 & 0.3 & --- \\
    ~~~- nonlinearity                 & --- & --- & 0.1 & 0.1 & --- \\
    Jet    identification efficiency  & --- & --- & 2.0 & 2.0 & ---  \\
    b jet  identification efficiency  & --- & --- & 2.9 & 3.5 & ---  \\
    Higgs boson background            &    &    &     &     & \\
    ~~~~- cross section scale factors & --- & --- & 9.3 & 9.3 & ---\\
    ~~~~- PDF                         & --- & --- & 8.1 & 8.1 & ---\\
    \hline
    b jet CSV distribution              &    &    &    &    & \\
    ~~~~- purity                        & --- & --- & --- & --- & 1.0--3.4\\
    ~~~~- statistical precision         & --- & --- & --- & --- & 1.0--24\phantom{.}\\
    \ttbar + heavy flavor jets          & --- & --- & --- & --- & 0.3--1.0\\
    Modeling W decay daughters          & --- & --- & --- & --- & 1.6--2.7\\
    Generator parameters                &    &    &    &    & \\
    ~~~- QCD scale                      & --- & --- & --- & --- & 1.0--4.9 \\
    ~~~- matching parton-jet  threshold & --- & --- & --- & --- & 1.3--9.4 \\
    ~~~- top quark mass                 & --- & --- & --- & --- & 0.8--4.1 \\
    \end{tabular}
    }
    \end{center}}
\end{table}

\section{Results}
\label{sec:results}

The expected number of events from the SM background processes and the
expected number of signal events in data assuming a branching fraction
$\mathcal{B}(\cPqt\to\PH\cPq) = 1\%$ are shown in
Tables~\ref{table:yieldsa}, \ref{tab:gg-s-bg}, and
\ref{table:multijet_fit_results} for the multilepton, diphoton, and b jet +
lepton selections, respectively.  The final results are based  on the
combination of 12 channels: three SS dilepton, four trilepton, one
diphoton + hadrons, two diphoton + lepton, and two b jet + lepton.  The
combination requires the simultaneous fit of the data selected by all
the individual analyses, accounting for all statistical and systematic
uncertainties, and their correlations.  As $\mathcal{B}({\rm t \to Hq})$
is expected to be small, the possibility of both top quarks decaying via
FCNC is not considered.

No excess beyond the expected SM background is observed and upper limits
at the 95\% CL on the branching fractions of $\rm t \to Hc$ and $\rm t
\to Hu$ are determined using the modified frequentist approach
(asymptotic CLs
method~\cite{ref:Junk:1999kv,ref:Read:2002hq,Cowan:2010js}).
The observed 95\% CL upper limits on the branching fractions ${\cal
B}({\rm t \to Hc})$ and ${\cal B}({\rm t \to Hu})$ are 0.40\% and
0.55\%, respectively, obtained from the combined multilepton, diphoton,
and b jet + lepton channels.  A summary
of the observed and expected limits is presented in
Table~\ref{table:limits_final}.  The diphoton channels are significantly
more sensitive than the other channels, largely because of the lower
uncertainty in the background model.  The multilepton and b jet + lepton
channels provide a 15\% (37\%) improvement on the observed (expected)
upper limit when combined with the diphoton channel.  A previous search
for FCNC mediated by Higgs boson interactions via the $\rm t \to Hc$ decay
at the LHC made use of trilepton and diphoton final
states~\cite{Khachatryan:2014jya}.  The inclusion of new channels (SS
dilepton, diphoton, and b jet + lepton final states) in addition to
refinements in the trilepton and diphoton channels results in an
improvement of 30\% (34\%) in the observed (expected) upper limit
on $\mathcal{B}(\rm t\to Hc)$.

The partial width of the ${\rm t \to Hq}$ process is related to the
square of the Yukawa coupling $\lambda_{\rm tq}$
by the formula~\cite{Craig:2012vj,Atwood}:

\begin{equation*}
    \Gamma_{\cPqt\to\PH\cPq}= \frac{m_{\cPqt}}{16\pi}
    \left|\lambda_{\rm tq}^{\PH}\right|^{2}
    \left[(y_{\rm q}+1)^2-y^2\right]
    \sqrt{1-(y-y_{\rm q})^2} \sqrt{1-(y+y_{\rm q})^2},
\end{equation*}

where $y = m_{\PH}/m_{\rm t}$ and $y_{\rm q} = m_{\rm q}/m_{\rm t}$.
(Note that a convention where the parity of the coupling is ignored is
adopted here: this introduces a factor of two when comparing to the
ATLAS result.) Assuming the ${\rm t \to Wb}$ partial width to be
dominant, the upper limit on the ${\rm t \to Hq}$ branching fractions
can be translated into an upper limit on the square of the couplings
using the relations:

\begin{equation*}
    \mathcal{B}({\rm t \to Hc}) = \Gamma_{\rm t\to Hc}/\Gamma_{\rm Total} =
    (0.58 \pm 0.01) \left|\lambda_{\rm tc}^{\PH}\right|^{2},
\end{equation*}

\begin{equation*}
    \mathcal{B}({\rm t \to Hu}) = \Gamma_{\rm t\to Hu}/\Gamma_{\rm Total} =
    (0.56 \pm 0.01) \left|\lambda_{\rm tu}^{\PH}\right|^{2},
\end{equation*}

where the CKM matrix element $\left|V_{\rm tb}\right|$ is assumed to be
equal to unity in the NLO order calculation~\cite{Denner} of
$\Gamma_{\rm Total}\approx\Gamma_{\rm t\to Wb}=1.372\GeV$, and
uncertanties arise from uncertainties on the mass values.  The Particle
Data Group~\cite{PDG} values of $m_{\rm H} = 125\GeV$, $m_{\rm t} =
173.5\GeV$, $m_{\rm c} = 1.29\GeV$, and $m_{\rm u}=2.3\MeV$ are used.

Based on the analysis results, the observed (expected) 95\% CL upper
limits on the squares of the top-Higgs Yukawa couplings are:

\begin{equation*}
    \left|\lambda_{\rm tc}^{\PH}\right|^{2} < 6.9~(7.4^{+3.6}_{-2.2}) \times 10^{-3},\\
\end{equation*}

\begin{equation*}
    \left|\lambda_{\rm tu}^{\PH}\right|^{2} < 9.8~(7.1^{+3.2}_{-2.3}) \times 10^{-3}.
\end{equation*}

\begin{table}[ht!]
    \topcaption{The observed and expected upper limits at
                the $95\%$ CL on the branching fraction (in \%) of $\rm t
                \to Hq$ (q = u, c) for: trilepton, SS dilepton, and
                combined multilepton channels; diphoton; b jet + lepton; and the
                combination of all channels.  For the expected upper limit, the
                limit plus and minus a standard deviation are also shown.
                } \label{table:limits_final}
    \vspace{0.1in}
    \centering
    \begin{tabular}{ l | c | c c c }
    & $\mathcal{B}_\text{obs}$($ \PQt \to \PH \PQc $) & $\mathcal{B}_\text{exp}$($ \PQt \to \PH \PQc $) & $\mathcal{B}_\text{exp}{+}\sigma$ & $\mathcal{B}_\text{exp}{-}\sigma$ \\
    \hline
    Trilepton            & 1.26 & 1.33 & 1.87 & 0.95 \\
    Same-sign dilepton   & 0.99 & 0.93 & 1.26 & 0.68 \\
    Multilepton combined & 0.93 & 0.89 & 1.22 & 0.65 \\
    Diphoton hadronic    & 1.26 & 1.33 & 1.87 & 0.95 \\
    Diphoton leptonic    & 0.99 & 0.93 & 1.26 & 0.68 \\
    Diphoton combined    & 0.47 & 0.67 & 1.06 & 0.44 \\
    b jet + lepton       & 1.16 & 0.89 & 1.37 & 0.60 \\
    \hline
    Full combination     & 0.40 & 0.43 & 0.64 & 0.30 \\
    \hline \hline
    & $\mathcal{B}_\text{obs}$($ \PQt \to \PH \PQu $) & $\mathcal{B}_\text{exp}$($ \PQt \to \PH \PQu $) & $\mathcal{B}_\text{exp}{+}\sigma$ & $\mathcal{B}_\text{exp}{-}\sigma$ \\
    \hline
    Trilepton            & 1.34 & 1.47 & 2.09 & 1.05 \\
    Same-sign dilepton   & 0.93 & 0.85 & 1.16 & 0.62 \\
    Multilepton combined & 0.86 & 0.82 & 1.14 & 0.60 \\
    Diphoton hadronic    & 1.26 & 1.33 & 1.87 & 0.95 \\
    Diphoton leptonic    & 0.99 & 0.93 & 1.26 & 0.68 \\
    Diphoton combined    & 0.42 & 0.60 & 0.96 & 0.39 \\
    b jet + lepton       & 1.92 & 0.84 & 1.31 & 0.57 \\
    \hline
    Full combination     & 0.55 & 0.40 & 0.58 & 0.27 \\
    \end{tabular}
\end{table}

\section{Summary}
\label{sec:conclusions}

\par
A search for flavor-changing neutral currents in the decay of a top
quark to a charm or up quark and a Higgs boson based on $\sqrt{s} =
8\TeV$ proton-proton collisions has been presented.  Samples of
multilepton, diphoton, and b jet + lepton events were selected from data
recorded with the CMS detector, corresponding to an integrated
luminosity of 19.7\fbinv.  The topologies ${\rm pp \to \ttbar \to}$
${\rm Hq + Wb}$ events, where q = u, c and H is allowed to decay into
WW, ZZ, $\tau \tau$, $\gamma\gamma$, and \bbbar.  No excess of
events above the SM background is observed, and branching fractions of
${\cal B}({\rm t \to Hc})$ larger than 0.40\% and ${\cal B}({\rm t \to
Hu})$ larger than 0.55\% are excluded at the 95\% confidence level.
These observed upper limits on ${\cal B}({\rm t \to Hq})$ and the
corresponding constraints on the top quark flavor-changing Higgs boson
Yukawa couplings are amongst the most stringent measured to date.

\begin{acknowledgments}
\hyphenation{Bundes-ministerium Forschungs-gemeinschaft Forschungs-zentren Rachada-pisek}

We congratulate our colleagues in the CERN accelerator departments for
the excellent performance of the LHC and thank the technical and
administrative staffs at CERN and at other CMS institutes for their
contributions to the success of the CMS effort. In addition, we
gratefully acknowledge the computing centers and personnel of the
Worldwide LHC Computing Grid for delivering so effectively the computing
infrastructure essential to our analyses. Finally, we acknowledge the
enduring support for the construction and operation of the LHC and the
CMS detector provided by the following funding agencies: the Austrian
Federal Ministry of Science, Research and Economy and the Austrian
Science Fund; the Belgian Fonds de la Recherche Scientifique, and Fonds
voor Wetenschappelijk Onderzoek; the Brazilian Funding Agencies (CNPq,
CAPES, FAPERJ, and FAPESP); the Bulgarian Ministry of Education and
Science; CERN; the Chinese Academy of Sciences, Ministry of Science and
Technology, and National Natural Science Foundation of China; the
Colombian Funding Agency (COLCIENCIAS); the Croatian Ministry of
Science, Education and Sport, and the Croatian Science Foundation; the
Research Promotion Foundation, Cyprus; the Secretariat for Higher
Education, Science, Technology and Innovation, Ecuador; the Ministry of
Education and Research, Estonian Research Council via IUT23-4 and
IUT23-6 and European Regional Development Fund, Estonia; the Academy of
Finland, Finnish Ministry of Education and Culture, and Helsinki
Institute of Physics; the Institut National de Physique Nucl\'eaire et
de Physique des Particules~/~CNRS, and Commissariat \`a l'\'Energie
Atomique et aux \'Energies Alternatives~/~CEA, France; the
Bundesministerium f\"ur Bildung und Forschung, Deutsche
Forschungsgemeinschaft, and Helmholtz-Gemeinschaft Deutscher
Forschungszentren, Germany; the General Secretariat for Research and
Technology, Greece; the National Scientific Research Foundation, and
National Innovation Office, Hungary; the Department of Atomic Energy and
the Department of Science and Technology, India; the Institute for
Studies in Theoretical Physics and Mathematics, Iran; the Science
Foundation, Ireland; the Istituto Nazionale di Fisica Nucleare, Italy;
the Ministry of Science, ICT and Future Planning, and National Research
Foundation (NRF), Republic of Korea; the Lithuanian Academy of Sciences;
the Ministry of Education, and University of Malaya (Malaysia); the
Mexican Funding Agencies (BUAP, CINVESTAV, CONACYT, LNS, SEP, and
UASLP-FAI); the Ministry of Business, Innovation and Employment, New
Zealand; the Pakistan Atomic Energy Commission; the Ministry of Science
and Higher Education and the National Science Centre, Poland; the
Funda\c{c}\~ao para a Ci\^encia e a Tecnologia, Portugal; JINR, Dubna;
the Ministry of Education and Science of the Russian Federation, the
Federal Agency of Atomic Energy of the Russian Federation, Russian
Academy of Sciences, and the Russian Foundation for Basic Research; the
Ministry of Education, Science and Technological Development of Serbia;
the Secretar\'{\i}a de Estado de Investigaci\'on, Desarrollo e
Innovaci\'on and Programa Consolider-Ingenio 2010, Spain; the Swiss
Funding Agencies (ETH Board, ETH Zurich, PSI, SNF, UniZH, Canton Zurich,
and SER); the Ministry of Science and Technology, Taipei; the Thailand
Center of Excellence in Physics, the Institute for the Promotion of
Teaching Science and Technology of Thailand, Special Task Force for
Activating Research and the National Science and Technology Development
Agency of Thailand; the Scientific and Technical Research Council of
Turkey, and Turkish Atomic Energy Authority; the National Academy of
Sciences of Ukraine, and State Fund for Fundamental Researches, Ukraine;
the Science and Technology Facilities Council, UK; the US Department of
Energy, and the US National Science Foundation.

Individuals have received support from the Marie-Curie program and the
European Research Council and EPLANET (European Union); the Leventis
Foundation; the A. P. Sloan Foundation; the Alexander von Humboldt
Foundation; the Belgian Federal Science Policy Office; the Fonds pour la
Formation \`a la Recherche dans l'Industrie et dans l'Agriculture
(FRIA-Belgium); the Agentschap voor Innovatie door Wetenschap en
Technologie (IWT-Belgium); the Ministry of Education, Youth and Sports
(MEYS) of the Czech Republic; the Council of Science and Industrial
Research, India; the HOMING PLUS program of the Foundation for Polish
Science, cofinanced from European Union, Regional Development Fund, the
Mobility Plus program of the Ministry of Science and Higher Education,
the National Science Center (Poland), contracts Harmonia
2014/14/M/ST2/00428, Opus 2013/11/B/ST2/04202, 2014/13/B/ST2/02543 and
2014/15/B/ST2/03998, Sonata-bis 2012/07/E/ST2/01406; the Thalis and
Aristeia programs cofinanced by EU-ESF and the Greek NSRF; the National
Priorities Research Program by Qatar National Research Fund; the
Programa Clar\'in-COFUND del Principado de Asturias; the Rachadapisek
Sompot Fund for Postdoctoral Fellowship, Chulalongkorn University and
the Chulalongkorn Academic into Its 2nd Century Project Advancement
Project (Thailand); and the Welch Foundation, contract C-1845.

\end{acknowledgments}

\bibliography{auto_generated}
\cleardoublepage \appendix\section{The CMS Collaboration \label{app:collab}}\begin{sloppypar}\hyphenpenalty=5000\widowpenalty=500\clubpenalty=5000\textbf{Yerevan Physics Institute,  Yerevan,  Armenia}\\*[0pt]
V.~Khachatryan, A.M.~Sirunyan, A.~Tumasyan
\vskip\cmsinstskip
\textbf{Institut f\"{u}r Hochenergiephysik,  Wien,  Austria}\\*[0pt]
W.~Adam, E.~Asilar, T.~Bergauer, J.~Brandstetter, E.~Brondolin, M.~Dragicevic, J.~Er\"{o}, M.~Flechl, M.~Friedl, R.~Fr\"{u}hwirth\cmsAuthorMark{1}, V.M.~Ghete, C.~Hartl, N.~H\"{o}rmann, J.~Hrubec, M.~Jeitler\cmsAuthorMark{1}, A.~K\"{o}nig, I.~Kr\"{a}tschmer, D.~Liko, T.~Matsushita, I.~Mikulec, D.~Rabady, N.~Rad, B.~Rahbaran, H.~Rohringer, J.~Schieck\cmsAuthorMark{1}, J.~Strauss, W.~Treberer-Treberspurg, W.~Waltenberger, C.-E.~Wulz\cmsAuthorMark{1}
% ...
\vskip\cmsinstskip
% ...
1:~~Also at Vienna University of Technology, Vienna, Austria\\
% ...

\end{sloppypar}
\end{document}